\documentclass[12pt,a4paper]{article}
	
\usepackage{a4wide}
\usepackage{amsmath}
\usepackage{amssymb}
\usepackage{amsfonts}
\usepackage{epsfig}
\usepackage{exscale}
\usepackage{float}
\usepackage{bbm}
\usepackage{dirtytalk}
\usepackage[numbers,sort&compress]{natbib}
\usepackage{caption}
\usepackage{subcaption}
\usepackage[inkscapearea=page]{svg}
\usepackage{adjustbox}

\newcommand{\Z}{{\mathbb{Z}}}
\newcommand{\R}{{\mathbb{R}}}
\newcommand{\C}{{\mathbb{C}}}

\newcommand{\p}{\partial}
\newcommand{\bra}[1]{\langle #1 \lvert}
\newcommand{\ket}[1]{\lvert #1 \rangle}

\newcommand{\pqty}[1]{\left( #1 \right)}
\newcommand{\bqty}[1]{\left[ #1 \right]}
\newcommand{\abs}[1]{\left\lvert #1\right\rvert}

\newcommand{\expval}[1]{\langle #1 \rangle}

\newcommand {\dv}[3][ ]{
  \ifx #1 { }
    \frac{d #2}{d #3}
  \else
    \frac{d^{#1} #2}{d #3^{#1}}
  \fi
}
\newcommand {\pdv}[3][ ]{
  \ifx #1 { }
    \frac{\partial #2}{\partial #3}
  \else
    \frac{\partial^{#1} #2}{\partial #3^{#1}}
  \fi
}
\newcommand {\fdv}[3][ ]{
  \ifx #1 { }
    \frac{\delta #2}{\delta #3}
  \else
    \frac{\delta^{#1} #2}{\delta #3^{#1}}
  \fi
}

\renewcommand{\Im}{\operatorname{Im}}
\renewcommand{\Re}{\operatorname{Re}}

\makeatletter
\@addtoreset{equation}{section}
\makeatother

\title{Bouncing Wave Packets, Ehrenfest Theorem, \\
and Uncertainty Relation based upon a new \\
Concept for the Momentum of a Particle in a Box}

\author{I.\ Albrecht, J.\ Herrmann, A.\ Mariani, U.-J.\ Wiese, and V.\ Wyss \\ \\
Albert Einstein Center for Fundamental Physics \\
Institute for Theoretical Physics, Bern University \\
Sidlerstrasse 5, CH-3012 Bern, Switzerland \\ \\}

\begin{document} 

\maketitle

\vspace{-1cm}

\begin{abstract} \normalsize

For a particle in a box, the operator $-i\partial_x$ is not self-adjoint and thus does not qualify as the physical momentum. As a result, in general the Ehrenfest theorem is violated. Based upon a recently developed new concept for a self-adjoint momentum operator, we reconsider the theorem and find that it is now indeed satisfied for all physically admissible boundary conditions. We illustrate these results for bouncing wave packets which first spread, then shrink, and return to their original form after a certain revival time. We derive a very simple form of the general Heisenberg-Robertson-Schr\"odinger uncertainty relation and show that our construction also provides a physical interpretation for it. 

\end{abstract}

\newpage
 
\section{Introduction}

The particle in a box is a classic problem in quantum mechanics. In one of its typical formulations, an otherwise free quantum mechanical particle is confined to a finite $1$-dimensional interval $\Omega = [-\frac{L}{2},\frac{L}{2}] \subset \R$. Once one imposes appropriate boundary conditions, solving the eigenvalue equation for the Hamiltonian is a textbook exercise. However, the choice of boundary conditions is not as trivial as one may assume, and involves several subtleties with interesting physical consequences. 

In particular, while often in the physics literature one does not distinguish between Hermitean and self-adjoint operators, self-adjointness (and not Hermiticity alone) is essential to ensure that an operator has a spectrum of real eigenvalues and a corresponding complete system of orthonormal eigenfunctions, at least in a generalized sense which includes distributions \cite{Gelfand}. In addition to Hermiticity (which results if an operator $A$ and its Hermitean conjugate $A^\dagger$ act in the same way), self-adjointness requires that the corresponding domains coincide, $D(A^\dagger) = D(A)$ \cite{VonNeumann32,Ree75}. 
The domain of an operator is usually characterized by square-integrability conditions on derivatives of the wave function as well as by boundary conditions, which are characterized by a family of self-adjoint extension parameters \cite{Bal70,Car90, AlHashimiWiese12}. The consistent interpretation of quantum mechanical measurements of an observable $A$, which returns one of its eigenvalues and collapses the wave function onto the corresponding eigenfunction, indeed requires that $A$ is self-adjoint \cite{VonNeumann32}. 

For the particle in a box, the operator $- i \p_x$ (here and throughout the paper we adopt units where $\hbar = 1$), which describes the momentum of a particle in the Hilbert space $L^2(\R)$ of square-integrable functions over the entire real line $\R$, is not self-adjoint in $L^2(\Omega)$ (at least if only local physical boundary conditions are imposed). This may be traced to the finite interval breaking translational invariance. For this reason, it has been concluded that momentum is no longer an observable in a finite interval \cite{Bon01}. Since the operator $- i \p_x$ is not self-adjoint in $L^2(\Omega)$, the problem is usually considered instead in $L^2(\R)$. Then the unquantized eigenvalues $k \in \R$ form a continuous spectrum. Since the corresponding eigenstates are plane waves $\exp(i k x)$, which exist everywhere in space with the same probability density, a momentum measurement of this type transfers an infinite amount of energy to the particle and catapults it out of the finite region.

While sharp impenetrable boundaries are a mathematical idealization, the physical model of the particle in a box and the new momentum concept may be applied as an effective description to many physical systems which are confined inside a limited region of space, such as ultra-cold atoms confined in an optical box trap \cite{UltraColdAtoms}, electrons in a quantum dot \cite{QuantumWellsWiresDots}, domain wall fermions in a higher-dimensional space \cite{Kap92, Sha93} or the phenomenological MIT bag model \cite{MITBagModel1,MITBagModel2, MITBagModel3} for confined quarks and gluons.

Recently, a self-adjoint momentum operator was constructed for a particle that is strictly confined to a finite 1-d interval 
$\Omega = [-\tfrac{L}{2},\tfrac{L}{2}]$, even after a momentum measurement \cite{alHashimiWieseAltMomentum,alHashimiWieseHalfLine}. 
In that case, the momentum eigenvalues are quantized. The key to this construction is the doubling of the Hilbert space to $L^2(\Omega) \times \C^2$, which was originally motivated by an ultraviolet lattice regularization, and led to a resolution of this long-standing puzzle also directly in the continuum.

In Section \ref{sec:postintro} we first provide an introduction to the problem and an outline of the new concept of momentum. We then explore several issues related to our construction. We take the point of view that boundary conditions for the particle, rather than being imposed by hand, should be derived from the self-adjointness requirement of relevant operators. This leads directly (in Section \ref{sec:wavefunctions})  to studying the spectrum of the particle in a box with the most general boundary conditions that make the Hamiltonian self-adjoint, displaying several features that are absent in the usual treatment.
In Section \ref{sec:ehrenfest}, we show that our construction of the self-adjoint momentum operator indeed satisfies the Ehrenfest theorem. This is not true in general for the usual momentum. In Section \ref{sec:bouncing} we then reconsider the classic problem of the time evolution of a Gaussian wave packet bouncing off the impenetrable boundary using a technique to wrap solutions on the infinite real line to the finite interval. In Section \ref{sec:uncertainty} we are able to provide a satisfactory interpretation of the finite-volume variant of the Heisenberg uncertainty relation, generalised by Robertson and Schr\"odinger. Finally, Section \ref{sec:conclusions} contains our conclusions.

\section{Boundary Conditions and the New Momentum Concept}\label{sec:postintro}

In the present section we first carefully derive the most general local boundary conditions which make the Hamiltonian self-adjoint. These turn out to depend on two real self-adjoint extension parameters $\gamma_{\pm}$. We consider the energy spectrum for some choices of $\gamma_\pm$, which shows some interesting features. We then present the new momentum concept and show some of its main features, including the spectrum and examples of measurements of the new momentum for some energy eigenstates.

\subsection{Self-adjoint Hamiltonians for a particle in a box}

Let us consider the Hamiltonian $H = - \tfrac{1}{2 m} \p_x^2 + V(x)$, $x \in \Omega = [-\tfrac{L}{2},\tfrac{L}{2}]$ (again we adopt units where $\hbar = 1$). We assume that $V(x)$ is a non-singular potential. We now illustrate the distinction between Hermiticity and self-adjointness for the Hamiltonian $H$. Performing two partial integrations, one obtains
\begin{align}
    \langle H^\dagger \chi|\Psi\rangle &= \langle\chi|H \Psi\rangle \nonumber =\\ 
    &=\langle H \chi|\Psi\rangle + \frac{1}{2 m} \bqty{\p_x\chi^*(x) \Psi(x) - \chi^*(x) \p_x \Psi(x)}_{-L/2}^{L/2} \ .
    \label{HHermiticity}
\end{align}
Hermiticity of $H$ requires the boundary term to vanish. The most general boundary conditions that ensure Hermiticity of $H$ while preserving the linearity of quantum mechanics as well as locality are Robin boundary conditions
\begin{equation}
    \gamma_{\pm} \Psi(\pm L/2) \pm \p_x \Psi(\pm L/2)=0 \ ,
    \label{Robinbc}
\end{equation}
with $\gamma_{\pm} \in \C$. Dirichlet boundary conditions, $\Psi(\pm L/2) = 0$, correspond to $\gamma_{\pm} \to \infty$, and Neumann boundary conditions, $\p_x \Psi(\pm L/2) = 0$, correspond to $\gamma_{\pm} = 0$.  Wave functions that obey eq.\eqref{Robinbc} and whose second derivative is square-integrable belong to the domain $D(H)$. Note that $\Psi$ and $\chi$ may in principle satisfy different boundary conditions, which correspond to the domains $D(H)$ and $D(H^\dagger)$. In fact, if $\Psi$ satisfies eq.\eqref{Robinbc}, the Hermiticity condition requires
\begin{equation}
    \gamma_{\pm}^* \chi(\pm L/2) \pm \p_x \chi(\pm L/2)=0 \ .
\end{equation}
The domains $D(H^\dagger)$ and $D(H)$ therefore coincide only if  $\gamma_{\pm}^* = \gamma_{\pm} \in \R$. This defines a $2$-parameter family of self-adjoint extensions of $H$, based on each possible choice of extension parameters $\gamma_{\pm} \in \R$. The boundary conditions of eq.\eqref{Robinbc} 
ensure that the probability current
\begin{equation}
    j(x) = \frac{1}{2 m i} [\Psi(x)^* \p_x \Psi(x) - \p_x \Psi(x)^* \Psi(x)] \ ,
    \label{probability current}
\end{equation}
does not flow outside the box, $j(\pm L/2)=0$. Self-adjointness is hence directly responsible for probability conservation.

\subsection{Wave functions and energy spectrum} \label{sec:wavefunctions}

The requirement of self-adjointness for the Hamiltonian has led us to consider Robin boundary conditions eq.\eqref{Robinbc}. In this section we set $V(x)\equiv 0$ inside $[-\frac{L}{2}, \frac{L}{2}]$ and discuss the resulting eigenfunctions and energy levels for various choices of self-adjoint extension parameters $\gamma_{\pm}$.

\subsubsection{Dirichlet or Neumann boundary conditions}

First we recall the result for Dirichlet or Neumann boundary conditions. In both cases the energy spectrum is non-negative. In fact, for the standard textbook case of Dirichlet boundary conditions, i.e.\ $\gamma_+ = \gamma_- = \infty$, the energy eigenvalues are given by
\begin{equation}
    E_l = \frac{\pi^2 (l + 1)^2}{2 m L^2} \ , \quad l \in \{0,1,2,\dots\} \ , 
    \label{dirichlet spectrum}
\end{equation}
so the spectrum is strictly positive and the corresponding eigenstates take the form
\begin{align}
    \expval{x|\psi_l} &= \sqrt{\frac{2}{L}} \cos\left(\frac{\pi (l + 1)}{L} x\right) \ , \quad l \ \mbox{even} \ , \nonumber \\
    \expval{x|\psi_l} &= \sqrt{\frac{2}{L}} \sin\left(\frac{\pi (l + 1)}{L} x\right) \ , \quad l \ \mbox{odd} \ .
\label{dirichlet eigenstates}
\end{align}

For Neumann boundary conditions  with $\gamma_+ = \gamma_- = 0$, on the other hand, we also have a zero-energy state. In fact, the energy eigenvalues take the form 
\begin{equation}
    E_l = \frac{\pi^2 l^2}{2 m L^2} \ , \quad l \in \{0,1,2,\dots\} \ ,
\end{equation}
and the energy eigenstates are given by
\begin{align}
    \expval{x|\psi_0} &= \frac{1}{\sqrt{L}} \ , \nonumber \\
    \expval{x|\psi_l} &= \sqrt{\frac{2}{L}} \cos\left(\frac{\pi l}{L} x\right) \ , \quad l \neq 0 \ \mbox{even} \ , \nonumber \\
    \expval{x|\psi_l} &= \sqrt{\frac{2}{L}} \sin\left(\frac{\pi l}{L} x\right)  \ , \quad l \ \mbox{odd} \ .
    \label{neumann eigenstates}
\end{align}

Let us also consider mixed Neumann-Dirichlet boundary conditions with $\gamma_- = 0, \gamma_+ = \infty$. The energy eigenvalues are then given by
\begin{equation}
E_l = \frac{\pi^2 (l + \tfrac{1}{2})^2}{2 m L^2} \ , \quad 
l \in \{0,1,2,\dots\} \ , 
\end{equation}
and the energy eigenstates are
\begin{equation}
\expval{x|\psi_l} = \sqrt{\frac{2}{L}} \sin\left(\frac{\pi (l + \tfrac{1}{2})}{L} (x - \tfrac{L}{2})\right) \ .
\end{equation}
This should be compared with both \eqref{dirichlet eigenstates} and \eqref{neumann eigenstates}.

\begin{figure}
    \centering
    \includegraphics[width=14cm]{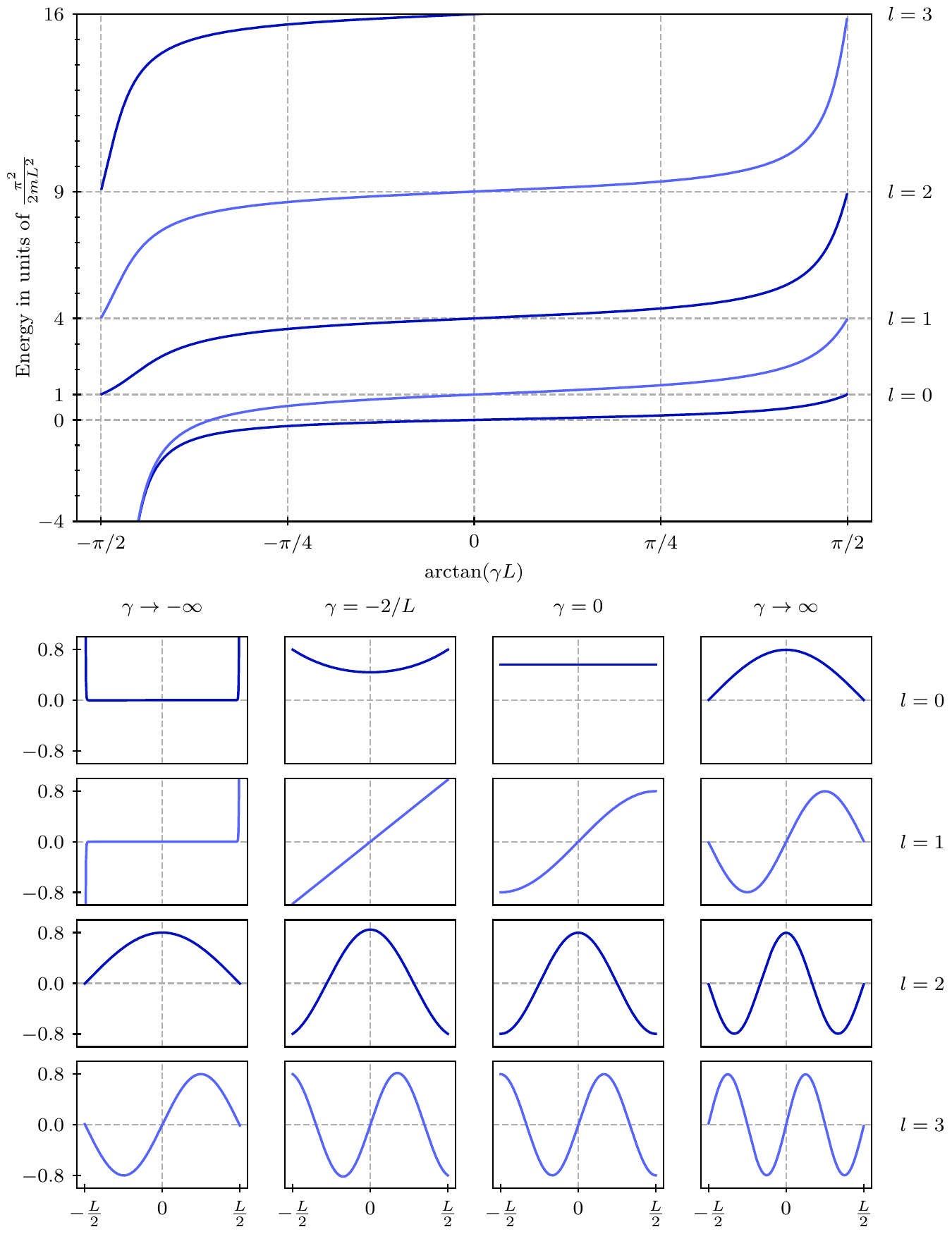}
    \caption{\textit{Top}: the energy of the first four, and part of the fifth, energy eigenstates in the parity-symmetric case $\gamma_+=\gamma_-\equiv\gamma$ as a function of $\arctan{\pqty{\gamma L}}$. \textit{Bottom}: the corresponding first four energy eigenstates at specific values of $\gamma$. }
    \label{fig:spectrum symmetric}
\end{figure}

\subsubsection{Parity-symmetric Hamiltonian}

In the parity-symmetric case, which has already been discussed in \cite{AlHashimiWieseUncertainty}, one has $\gamma_+=\gamma_-\equiv\gamma$. Note that both pure Dirichlet and pure Neumann boundary conditions are parity-symmetric, but mixed Neumann-Dirichlet boundary conditions are not.

In this case we have, first of all, a discrete positive energy spectrum with energies $E_l = \frac{k_l^2}{2m}$, where the $k_l$ are implicitly defined by
\begin{align}
    k_l \tan\pqty{k_l \tfrac{L}{2}} &= \gamma, \quad\quad l = 0,2,4,\ldots\nonumber\\
    k_l \cot\pqty{k_l \tfrac{L}{2}} &= -\gamma, \quad\quad l = 1,3,5,\ldots
\end{align}
Each of these two equations admit a countable infinity of solutions, ordered such that $k_l < k_{l'}$ for $l < l'$. The corresponding eigenstates are
\begin{equation}
    \expval{x|\psi_l} = \sqrt{\frac{2}{L}}\begin{cases} \pqty{1+\frac{\sin{(k_l L)}}{k_l L}}^{-1/2} \cos{(k_l x)} \ , & l = 0,2,4,\ldots\\ \pqty{1-\frac{\sin{(k_l L)}}{k_l L}}^{-1/2} \sin{(k_l x)}\ , & l = 1,3,5,\ldots\end{cases}
\end{equation}
These all have definite parity as expected. At specific values of $\gamma$, as we saw in the case of Neumann boundary conditions, we also have zero-energy states,
\begin{align}
    \expval{x|\psi_0} &= \frac{1}{\sqrt{L}} \ , \quad\quad \mathrm{for}\,\, \gamma = 0 \ ,\nonumber\\
    \expval{x|\psi_1} &=\sqrt{\frac{12}{L^3}} x \ , \quad\quad \mathrm{for}\,\, \gamma = -\frac{2}{L} \ .
    \label{zero-energy states symmetric}
\end{align}
Interestingly, in this case we may also have negative energy states. Setting $E = - \frac{\varkappa^2}{2m}$ we now must have
\begin{align}
    \varkappa \tanh(\varkappa \tfrac{L}{2}) &= -\gamma \ ,\nonumber\\
    \varkappa \coth(\varkappa \tfrac{L}{2}) &= -\gamma \ .
\end{align}
Up to the trivial reflection $\varkappa \to -\varkappa$, the first equation admits a unique solution for $\gamma < 0$ and no solution otherwise, while the second equation admits a unique solution for $\gamma < -2/L$ and no solution otherwise. Thus, depending on the value of $\gamma$, we have either one, two, or no negative energy solutions. The corresponding eigenstates are given by
\begin{align}
    \expval{x|\psi_0} &= \sqrt{\frac{2}{L}}\pqty{\frac{\sinh{(\varkappa L)}}{\varkappa L}+1}^{-1/2} \cosh{(\varkappa x)} \ ,\nonumber\\
    \expval{x|\psi_1} &= \sqrt{\frac{2}{L}}\pqty{\frac{\sinh{(\varkappa L)}}{\varkappa L}-1}^{-1/2} \sinh{(\varkappa x)} \ .
\end{align}
The dependence on $\gamma$ of the first four energy levels and their wave functions is depicted in Fig.\ref{fig:spectrum symmetric}. In fact, as one lowers $\gamma$, the first two energy levels for Dirichlet boundary conditions ($\gamma=\infty$) become the first two energy levels for Neumann boundary conditions ($\gamma=0$) and then, at the specific values of $\gamma$ derived above, they become the two zero-energy states of eq.\eqref{zero-energy states symmetric}. As $\gamma \to -\infty$, one might again expect a Dirichlet-like spectrum, and this is indeed true for the positive energy states. However, the energy of the two former zero-energy states now diverges to $-\infty$. As is shown in Fig.\ref{fig:spectrum symmetric}, the wavefunction of these two states is localized at the boundary in the limit $\gamma \to -\infty$.

\subsubsection{Hamiltonian with $\gamma_+=-\gamma_-$}\label{sec:antisymmetric hamiltonian}

\begin{figure}
    \centering
    \includegraphics[width=14cm]{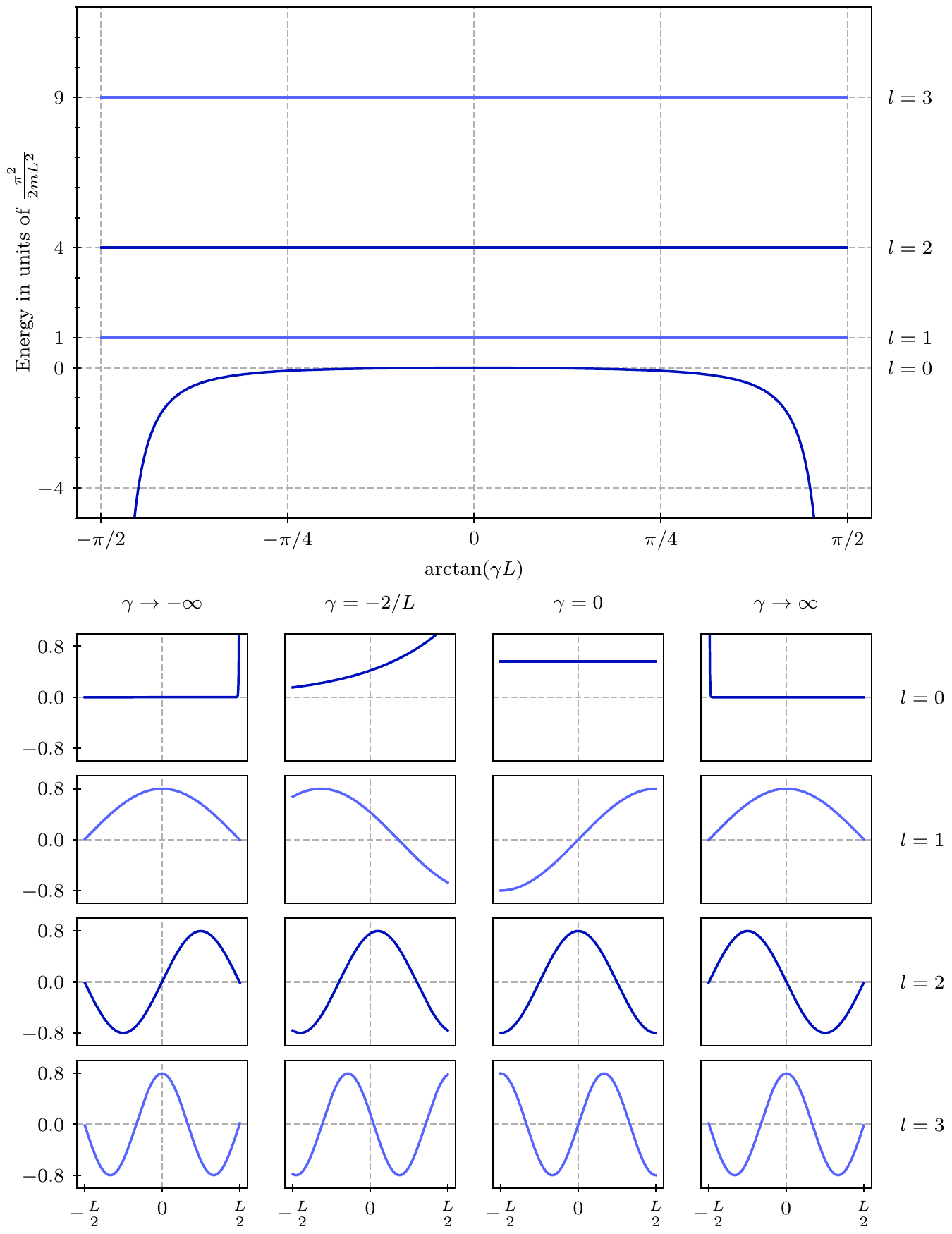}
    \caption{\textit{Top}: the energy of the first four energy eigenstates in the case $\gamma_+=-\gamma_-\equiv\gamma$ as a function of $\arctan{\pqty{\gamma L}}$. \textit{Bottom}: the corresponding first four energy eigenstates at specific values of $\gamma$. }
    \label{fig:spectrum antisymmetric}
\end{figure}

In this case we choose $\gamma_+=-\gamma_-\equiv\gamma$. The spectrum is drastically different from the previous case. The positive energy spectrum is given by
\begin{equation}
    E_l = \frac{k_l^2}{2m} \ ,\quad \quad \quad k_l = \frac{\pi}{L}l, \quad l = 1,2,3,\ldots
\end{equation}
which is the same spectrum as for Dirichlet boundary conditions, eq.\eqref{dirichlet spectrum}. Strikingly, the positive energy spectrum is independent of $\gamma$. The eigenfunctions, however, do depend on $\gamma$:  
\begin{equation}
    \expval{x | \psi_l} = \frac{1}{\sqrt{2L(\gamma^2+k_l^2)}} \bqty{\pqty{\gamma-ik_l} \exp{\pqty{ik_l x}} - (-1)^l \pqty{\gamma+ik_l} \exp{\pqty{-ik_l x}}} \ .
    \label{positive eigenstates antisymmetric hamiltonian}
\end{equation}
It should be noted that the eigenstates \eqref{positive eigenstates antisymmetric hamiltonian} are real up to a constant phase factor. This fact is used in drawing the figure. 
There is only one zero-energy state given by $\expval{x|\psi_0} = \frac{1}{\sqrt{L}}$ at $\gamma = 0$. This is not surprising, as again for $\gamma=0$ the boundary conditions reduce to Neumann boundary conditions. For any real $\gamma$ there is a unique negative energy eigenstate with energy $E=-\frac{\gamma^2}{2m}$ and eigenfunction
\begin{equation}
    \label{eq:negative energy state}
    \psi(x) = \sqrt{\frac{\gamma}{\sinh(\gamma L)}} \exp{\pqty{-\gamma x}} \ .
\end{equation}
Remarkably, this is the only state in the spectrum whose energy depends on $\gamma$. The spectrum is summarized in Fig.\ref{fig:spectrum antisymmetric} for the first four eigenstates and the corresponding eigenfunctions.

\subsection{Standard momentum operator for a particle in 
a box}

While it is straightforward to construct self-adjoint extensions of the Hamiltonian, for the standard momentum operator $p = - i \p_x$ this is not possible in a physically meaningful way. We can attempt to derive appropriate boundary conditions as for the Hamiltonian. Again, via partial integration one has 
\begin{align}
\langle p^\dagger \chi|\Psi\rangle &= 
\langle\chi|p\Psi\rangle = \nonumber \\ 
&=\langle p \chi|\Psi\rangle - i \bqty{\chi(x)^*\Psi(x)}_{-L/2}^{L/2} \ .
\label{pHermiticity}
\end{align}
Hermiticity requires that the boundary term vanishes. For physical reasons we again limit ourselves to local boundary conditions, which do not relate the wave function values at physically distinct points. Hermiticity then results for Dirichlet boundary conditions, $\Psi(\pm L/2) = 0$, which define the domain $D(p)$. However, when $\Psi$ is fixed to zero at the boundary, $\chi$ can still take arbitrary values. As a consequence, the domain of $p^\dagger$ (which acts on $\chi(x)$) remains unrestricted and $D(p) \subset D(p^\dagger)$. Therefore with Dirichlet boundary conditions the operator $p=- i \p_x$ is Hermitean but not self-adjoint. This distinction has striking consequences of physical significance: in fact, $p=- i \p_x$ with Dirichlet boundary conditions has no eigenfunctions and it therefore cannot be measured in any reasonable sense. Consequently, it does not qualify as a physically acceptable momentum operator in the Hilbert space $L^2(\Omega)$.

If one relaxes the locality requirement on the boundary conditions, one can impose instead the linear, but non-local boundary condition
\begin{equation}
    \Psi(L/2)=\lambda \Psi(-L/2) \ ,
\end{equation}
where $\lambda \in \C$ is a parameter. This would be a natural choice if the two endpoints of the box would coincide, i.e.\ if the box was actually a circle. However, if the two endpoints are distinct, then it is unphysical to relate the wave functions at the two boundary endpoints. In any case, with these boundary conditions for $\Psi$, Hermiticity is guaranteed if $\chi$ satisfies
\begin{equation}
    \chi(L/2)=\frac{1}{\lambda^*} \chi(-L/2) \ .
\end{equation}
Now the domains $D(p)$ and $D(p^\dagger)$ coincide if $\lambda = \frac{1}{\lambda^*}$, that is if $\abs{\lambda}=1$. Therefore, for each choice of phase factor $\lambda$, one obtains a different self-adjoint momentum operator. As previously remarked, however, this construction requires unphysical non-local boundary conditions, and the question remains whether it is possible to define a self-adjoint momentum operator with local boundary conditions.

\subsection{A new concept for the momentum}

As pointed out recently \cite{alHashimiWieseAltMomentum, alHashimiWieseHalfLine}, a physically and mathematically satisfactory momentum operator can be defined in a doubled Hilbert space with 2-component wave functions
\begin{equation}
p_R = - i \left(\begin{array}{cc} 0 & \p_x \\ \p_x & 0 
\end{array}\right) = - i \sigma_1 \p_x, \
\Psi(x) = \left(\begin{array}{c} \Psi_e(x) \\ \Psi_o(x) 
\end{array}\right).
\label{2component}
\end{equation}
This choice is motivated by a lattice construction, which may be found in \cite{alHashimiWieseAltMomentum, alHashimiWieseHalfLine}. The notation $\Psi_{e,o}$ refers to even and odd lattice points, a distinction which survives in the continuum in the form of a doubled Hilbert space. Besides $p_R$, the new momentum operator $p = p_R + i p_I$ also has an anti-Hermitean contribution $i p_I$, which we will discuss later. In order to make this paper self-contained, for the benefit of the reader we review some aspects of \cite{alHashimiWieseAltMomentum, alHashimiWieseHalfLine}. First we will show that $p_R$ is self-adjoint. By partial integration one obtains
\begin{align}
\langle p_R^\dagger \chi|\Psi\rangle &= \langle \chi| p_R \Psi\rangle 
= \nonumber \\
&=\langle p_R \chi|\Psi\rangle -
i  \bqty{\chi_e(x)^* \Psi_o( x) + 
\chi_o(x)^* \Psi_e(x)}_{-L/2}^{L/2} \ .
\label{pRHermiticity}
\end{align}
The most general linear, local boundary condition that we can impose on $\Psi$ for the purpose of making $p_R$ Hermitean is therefore given by
\begin{equation}
    \Psi_o(\pm L/2) = \lambda_{\pm} \Psi_e(\pm L/2) \ ,
    \label{momentumbc}
\end{equation}
where $\lambda_\pm \in \C$ are two parameters. Inserting this in eq.\eqref{pRHermiticity}, Hermiticity of $p_R$ is then guaranteed by again imposing linear, local boundary conditions on $\chi$,
\begin{equation}
    \chi_o(\pm L/2) = -\lambda_{\pm}^* \chi_e(\pm L/2) \ .
\end{equation}
Self-adjointness of $p_R$ requires $D(p_R^\dagger) = D(p_R)$, 
which implies $\lambda_{\pm} = - \lambda_{\pm}^*$ such that 
$\lambda_{\pm} \in i \R$. Hence, there is a 2-parameter family of 
self-adjoint extensions, characterized by a purely imaginary parameter 
$\lambda_{\pm}$ at each of the two points on the boundary. 
The other component of the new momentum concept is the operator\footnote{Note that the preprint versions of \cite{alHashimiWieseAltMomentum, alHashimiWieseHalfLine} contain an incorrect extra factor of $1/2$ in the definition of $p_I$, which was corrected in the final published versions (e.g. see \cite{alHashimiWieseHalfLine} eq.(30)).}
\begin{equation}
    p_I = \lim_{\epsilon \to 0} \begin{pmatrix} \delta(x+L/2-\epsilon)-\delta(x-L/2+\epsilon) & 0 \\ 0 & 0 \end{pmatrix} \ .
    \label{pI definition}
\end{equation}
The motivation for this choice again comes from a lattice construction, which will not be repeated here. The matrix elements of $p_I$ can be easily computed:
\begin{align}
    \bra{\chi}p_I\ket{\Psi} &= \lim_{\epsilon \to 0} \int_{-L/2}^{L/2} \chi_e^*(x)[\delta(x+L/2-\epsilon)-\delta(x-L/2+\epsilon)] \Psi_e(x) =\nonumber \\ 
    &=\bqty{\chi_e^*(-L/2)\Psi_e(-L/2)-\chi_e^*(L/2)\Psi_e(L/2)} \ .
    \label{pI matrix elements}
\end{align}
The operator $p_I$ is Hermitean and bounded, and thus naturally self-adjoint.

\subsubsection{Spectrum of $p_R$ for general boundary conditions}\label{sec:momentum spectrum}

For later convenience it is interesting to solve the eigenvalue problem for the new momentum concept $p_R$. The eigenvalue equation $p_R \phi_k(x) = k \phi_k(x)$ readily admits the general solution
\begin{equation}
    \phi_k(x) = \begin{pmatrix} A \exp{\pqty{ikx}} + B \exp{\pqty{-ikx}} \\ A \exp{\pqty{ikx}} - B \exp{\pqty{-ikx}} \end{pmatrix} \ .
    \label{pR eigenfunctions general solution}
\end{equation}
Applying the boundary condition eq.\eqref{momentumbc}, which makes the momentum self-adjoint, leads to the quantization condition
\begin{equation}
    \exp{\pqty{2ikL}} = \frac{(1+\lambda_+)(1-\lambda_-)}{(1-\lambda_+)(1+\lambda_-)} \equiv \exp{\pqty{2i\theta}} \ ,
    \label{pR quantization condition}
\end{equation}
for $\lambda_{\pm} \in i\R$ and the normalized eigenfunctions
\begin{equation}
    \phi_k(x) = \frac{1}{2\sqrt{L}}\begin{pmatrix}  \exp{\pqty{ikx}} + \sigma_k \exp{\pqty{-ikx}} \\ \exp{\pqty{ikx}} - \sigma_k \exp{\pqty{-ikx}} \end{pmatrix} \ , \quad \quad \sigma_k = \exp{\pqty{ikL}} \frac{1-\lambda_+}{1+\lambda_+} \ .
    \label{pR eigenfunctions normalized}
\end{equation}
The right-hand side of eq.\eqref{pR quantization condition} is a phase which depends only on $\lambda_{\pm}$; for convenience we called it $\exp{\pqty{2i\theta}}$ for some $\theta \in \R$. Then the solution of eq.\eqref{pR quantization condition} is given by
\begin{equation}
    k_n = \frac{\pi}{L}n + \frac{\theta}{L} \ .
    \label{pR eigenvalues}
\end{equation}
Therefore we find that the new momentum concept leads to the quantization of momentum for the particle in a box.

\subsubsection{The Hamiltonian and the physical Hilbert space}\label{sec:new hamiltonian}

We now need to reconsider the original Hamiltonian in light of the introduction of the doubled Hilbert space required to make sense of the new momentum operator. If one naively chooses the new Hamiltonian to be diagonal in the doubled space, i.e.\ $H' = \mathbbm{1} H$ where $H$ is the original Hamiltonian, then it will have eigenstates $\psi'_l(x) = \frac{1}{\sqrt{2}}\begin{pmatrix} \psi_l(x) \\ \pm \psi_l(x) \end{pmatrix}$ with energy $E_l$ where $\psi_l(x)$ and $E_l$ are eigenstates and eigenvalues of the original Hamiltonian. In other words, each energy level in the spectrum would become doubly degenerate. Thus the physics of the model would be different. For this reason, we need to modify the naive choice of the new Hamiltonian in such a way as to preserve the spectrum of the original theory. To this end, we note that each state $\Psi$ in the doubled Hilbert space may be split into the sum of two states, one with $\Psi_e=\Psi_o$ and the other one with $\Psi_e=-\Psi_o$:
\begin{equation}
    \langle x \lvert \Psi \rangle = \begin{pmatrix} \Psi_e(x) \\ \Psi_o(x) \end{pmatrix} = \frac{1}{2}\begin{pmatrix} \Psi_e(x)+\Psi_o(x) \\ \Psi_e(x) + \Psi_o(x) \end{pmatrix} + \frac{1}{2}\begin{pmatrix} \Psi_e(x)-\Psi_o(x) \\ -\Psi_e(x) + \Psi_o(x) \end{pmatrix} =\Psi^+(x) + \Psi^-(x) \ .
\end{equation}
We then modify the Hamiltonian to assign a different energy to the states with $\Psi_e=-\Psi_o$:
\begin{equation}
    H(\mu) = \begin{pmatrix} H & 0 \\ 0 & H \end{pmatrix} + \frac{\mu}{2} \begin{pmatrix} 1 & -1 \\ -1 & 1\end{pmatrix} = \mathbbm{1} H + \mu P_- \ ,
\end{equation}
where $P_-$ projects onto the subspace of the $\Psi^-$ and $\mu$ is an adjustable energy scale. Now the energy eigenstates and eigenfunctions are given in terms of the original eigenstates and eigenfunctions by
\begin{equation}
    H(\mu) \psi_l^+ = E_l \psi_l^+ \ ,\quad\quad\quad H(\mu) \psi_l^- = \pqty{E_l+\mu} \psi_l^- \ ,
\end{equation}
where $\psi_l^{\pm}(x)$ are the projections onto two subspaces of the eigenfunctions of the original Hamiltonian. Thus in the limit $\mu\to\infty$ only the $\Psi^+$ states have finite energy, while the $\Psi^-$ states are removed from the spectrum, which is now identical to the one of the original Hamiltonian. The origin of this choice of projection rests on the original lattice motivation for the new momentum concept \cite{alHashimiWieseAltMomentum, alHashimiWieseHalfLine}. In the lattice formulation, one finds that the $\Psi^-$ states have energies of the order of the inverse lattice spacing, i.e.\ of the order of the energy cut-off. These states then have infinite energy in the continuum limit. Thus only those states in the doubled Hilbert space that obey $\Psi_e = \Psi_o$ are considered as belonging to the physical, finite-energy Hilbert space. 

It is important to note that while the physical finite-energy states obey $\Psi_e = \Psi_o$, in order to properly describe momentum we still need the entire doubled Hilbert space. In fact, no state in the domain of $p_R$ (eq.\eqref{pR eigenfunctions normalized}) satisfies $\Psi_e\pqty{\pm\tfrac{L}{2}} = \Psi_o\pqty{\pm\tfrac{L}{2}}$. unless $\Psi_e\pqty{\pm\tfrac{L}{2}} = \Psi_o\pqty{\pm\tfrac{L}{2}}=0$. In particular, this means that one cannot naively apply $p_R$ to a state in the finite-energy subspace. However, since $p_R$ is self-adjoint its eigenstates form a basis for the entire Hilbert space and we can therefore use the spectral decomposition of $p_R$ to compute relevant quantities for finite-energy states. We will use this fact in Section \ref{sec:ehrenfest}.

One can then ask what boundary conditions make the new Hamiltonian with finite $\mu$ self-adjoint. Since we want the physics of the model to reproduce that of the original Hamiltonian, the self-adjoint extension parameters must allow for $\Psi_e\pqty{\pm\tfrac{L}{2}}=\Psi_o\pqty{\pm\tfrac{L}{2}}$. This then results in Robin boundary conditions on the finite-energy sector $\Psi^+$ \cite{alHashimiWieseAltMomentum}.


\subsubsection{Momentum measurements} \label{sec:momentum measurements}

Because of the boundary conditions eq.\eqref{momentumbc}, unless $\gamma_+=\gamma_-=\infty$, no state in the domain of $p_R$ satisfies $\Psi_e\pqty{\pm\tfrac{L}{2}}=\Psi_o\pqty{\pm\tfrac{L}{2}}$ unless $\Psi_e\pqty{\pm\tfrac{L}{2}}=\Psi_o\pqty{\pm\tfrac{L}{2}}=0$, and therefore in general no eigenstate of $p_R$ lies in the finite-energy sector. This means that every momentum measurement on a state in the finite-energy sector will lead outside the finite-energy sector, and thus an idealized momentum measurement leads to the transfer of infinite energy to the particle. For this reason, while one can use the momentum $p_R$ to compute the probability of measuring different momentum eigenvalues, what happens after a momentum measurement is not captured by the effective model of the particle in a box and will depend on the underlying ultraviolet details \cite{alHashimiWieseAltMomentum, alHashimiWieseHalfLine}. In other words, the particle in a box may be regarded as an effective description of a physical system confined in a finite region of space; however, the sharp impenetrable boundaries give rise to a high degree of ultraviolet sensitivity, with the fine ultraviolet details not captured by this effective description.

\begin{figure}
    \centering
    \begin{subfigure}[b]{0.9\textwidth}
        \centering
        \adjustbox{trim=0 0 0 {.11\height},clip}{\includegraphics[width=12cm]{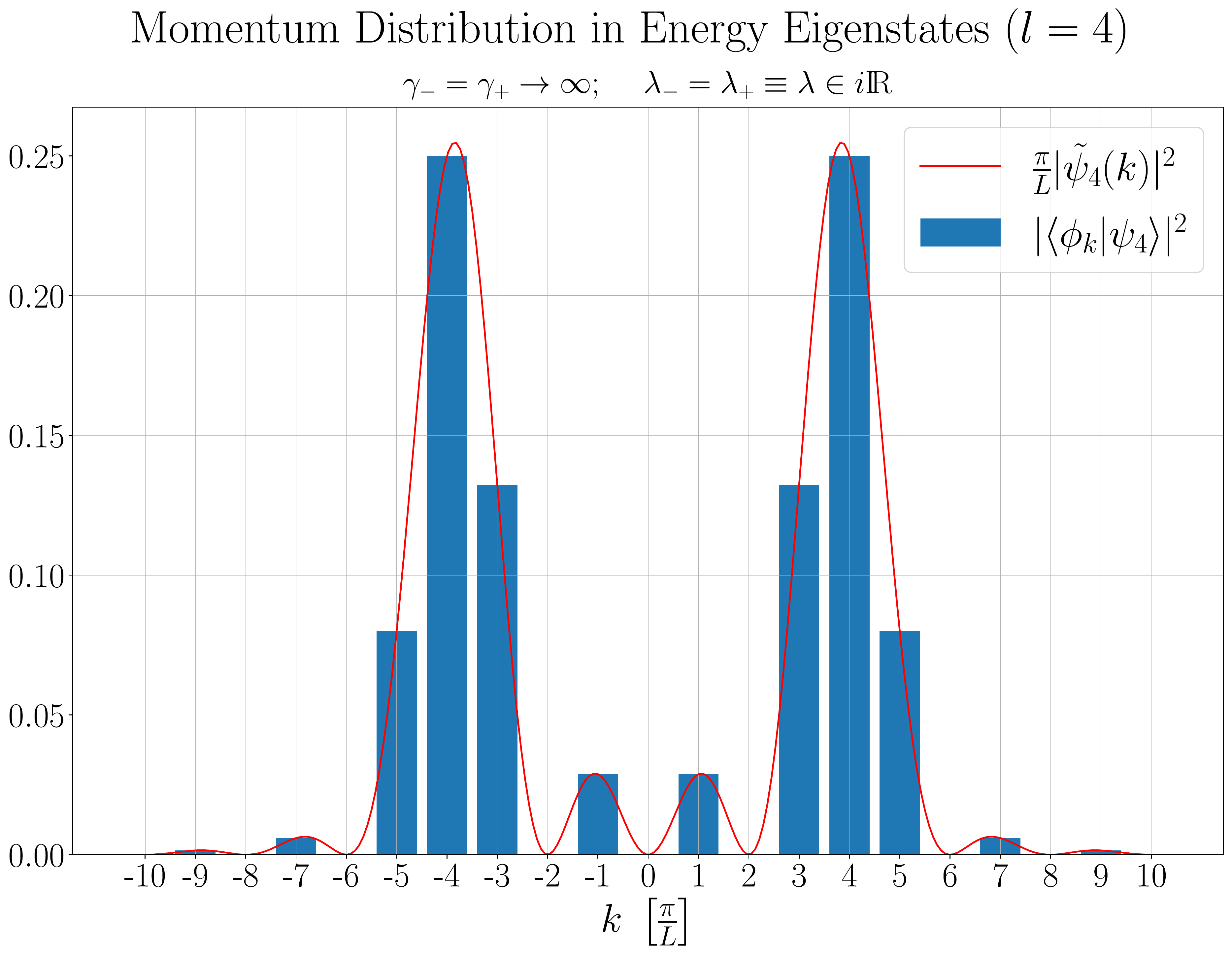}}
        \caption{$l=4$}
    \end{subfigure}
    \hfill
    \begin{subfigure}[b]{0.9\textwidth}
        \centering
        \adjustbox{trim=0 0 0 {.11\height},clip}{\includegraphics[width=12cm]{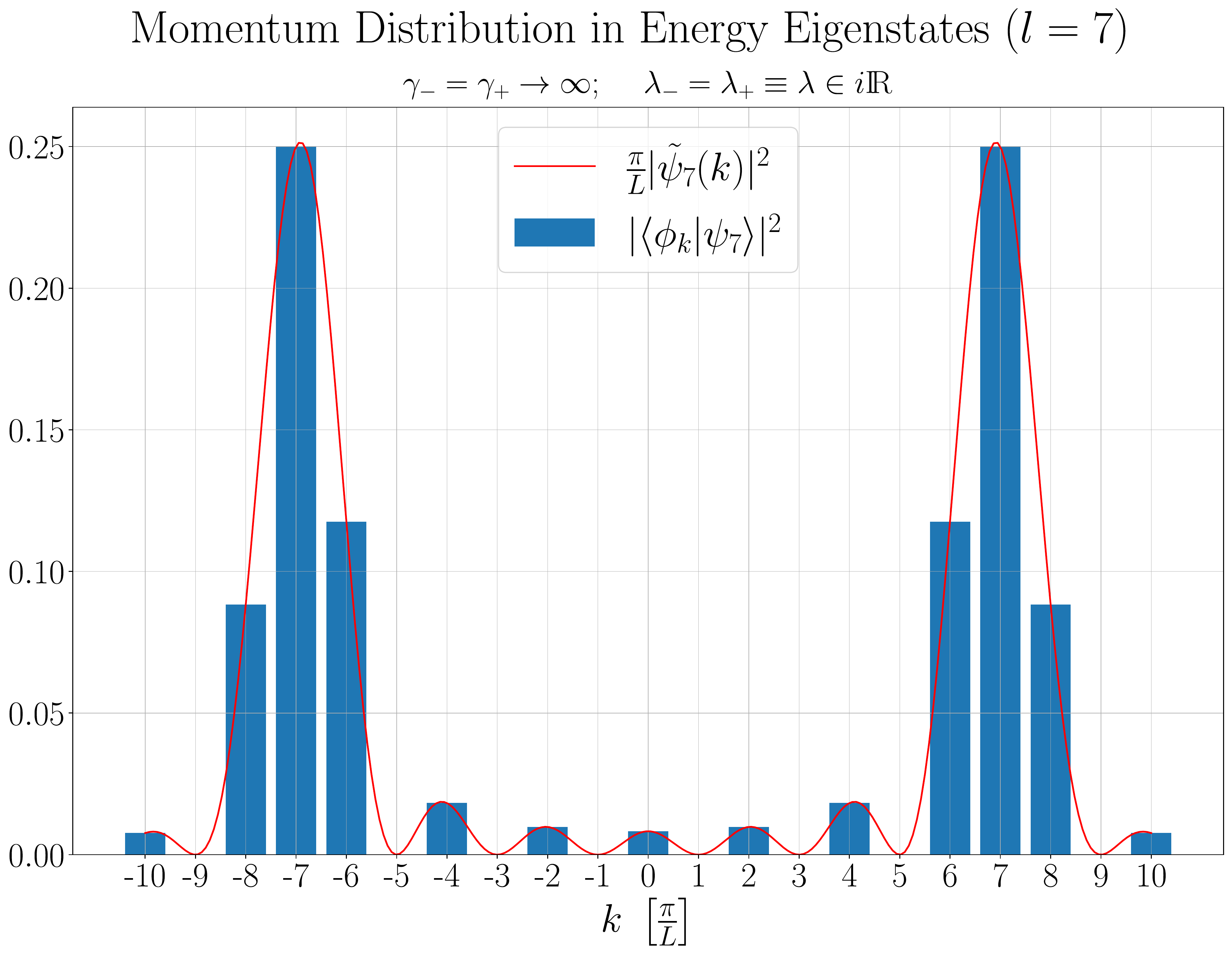}}
        \caption{$l=7$}
    \end{subfigure}
    \hfill
    \caption{The histogram shows the probability of measuring the eigenvalues $k_n=\tfrac{\pi}{L}n$ of the new momentum $p_R$ in the $l=4$ and $l=7$ energy eigenstates with Dirichlet boundary conditions. The red line is proportional to the probability density to measure the unquantized values of the standard momentum \cite{alHashimiWieseAltMomentum,alHashimiWieseHalfLine}.}
    \label{fig:momentum measurement dirichlet}
\end{figure}

After measuring the momentum $p_R$, the particle will end up in one of its eigenstates \eqref{pR eigenfunctions normalized}. Consider a generic normalized state in the finite-energy sector:
\begin{equation}
    \langle x \lvert \Psi \rangle = \frac{1}{\sqrt{2}} \begin{pmatrix} \Psi(x) \\ \Psi(x) \end{pmatrix} \ .
\end{equation}
The projection on a momentum eigenstate $\phi_k$ is given by
\begin{equation}
    \langle \phi_k \lvert \Psi \rangle = \frac{1}{\sqrt{2L}} \int_{-L/2}^{L/2} dx\, \exp{\pqty{-ik_n x}} \Psi(x) \equiv \sqrt{\frac{\pi}{L}} \widetilde{\Psi}(k_n) \ .
\end{equation}
The probability of measuring a momentum $k_n = \frac{\pi}{L}n + \frac{\theta}{L}$ is thus proportional to the modulus squared of the Fourier transform of the wave function at $k_n$. It is interesting to note that $\widetilde{\Psi}(k)$ is the probability amplitude to measure the unquantized value $k$ of the standard momentum \cite{alHashimiWieseAltMomentum,alHashimiWieseHalfLine}. It should be noted again that such a measurement would catapult the particle outside the box.

We now illustrate the results in a few cases of interest. Suppose that we insist on parity symmetry so that the momentum $p_R$ changes sign under parity. Then we must have $\lambda_+=\lambda_-$ which implies $\theta \equiv 0$. For an eigenstate $\psi_l$ of the Hamiltonian with Dirichlet boundary conditions, given in eq.\eqref{dirichlet eigenstates}, one finds
\begin{equation}
    \abs{\langle \phi_k \lvert \psi_l \rangle}^2 = \begin{cases} \frac14 & n = \pm l \ , \\
    0 & n+l \,\,\mathrm{even} \ , \\
    \frac{4 l^2}{\pi^2\pqty{l^2-n^2}^2} & n+l \,\,\mathrm{odd} \ , \end{cases}
    \label{dirichlet measurement}
\end{equation}
with $k_n = \frac{\pi}{L} n$. The probabilities add up to $1$ as expected, and the most probable $k$ is obtained when $n = \pm l$, i.e.\ when $k^2/2m = E_l$. This situation is illustrated for $l=4$ and $l=7$ in Fig.\ref{fig:momentum measurement dirichlet}. From eq.\eqref{dirichlet measurement} one may explicitly compute the expectation value of $p_R$ and its square,
\begin{equation}
    \expval{p_R} = 0 \ , \quad \quad \expval{p_R^2} = \frac{\pi^2 l^2 }{L^2} \ .
\end{equation}
In this case both expectation values are finite.

\begin{figure}
    \centering
    \adjustbox{trim=0 0 0 {.16\height},clip}{\includegraphics[width=12cm]{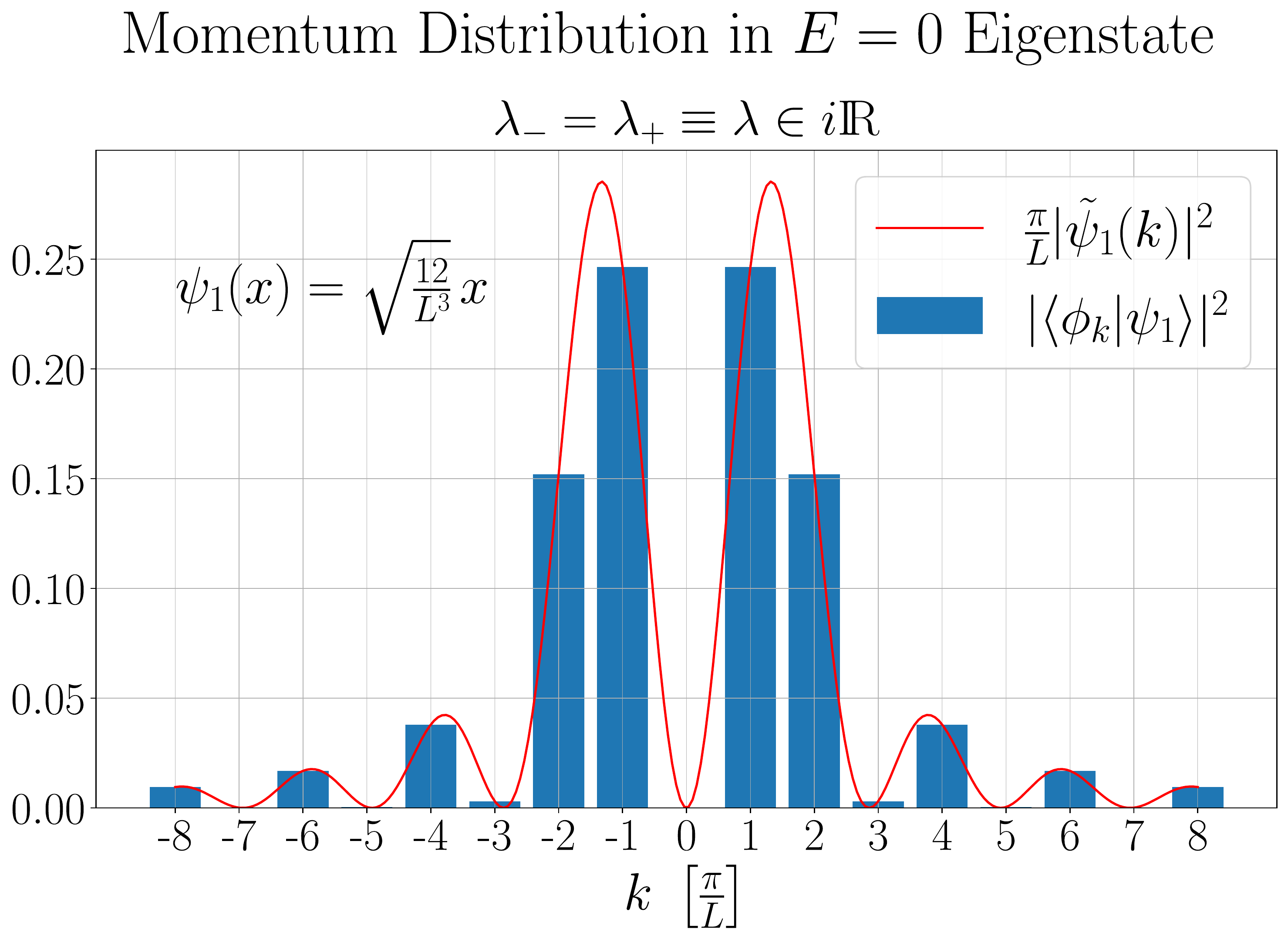}}
    \caption{Probability of measuring momentum $k_n=\tfrac{\pi}{L}n$ in the zero-energy eigenstate eq.\eqref{zero-energy states symmetric} with parity-symmetric boundary conditions.}
    \label{fig:momentum measurement zero-energy}
\end{figure}

To contrast this situation, we consider the linear zero-energy state \eqref{zero-energy states symmetric} obtained in the parity-symmetric case with $\gamma_+=\gamma_-=-2/L$. Here we get
\begin{equation}
    \label{zero-energy measurement}
    \abs{\langle \phi_k \lvert \psi_1 \rangle}^2 = \begin{cases} 0 & n = 0 \ , \\
    \frac{6}{\pi^2 n^2} & n \neq 0 \,\,\mathrm{even} \ , \\ \frac{24}{\pi^4 n^4} & n\,\,\mathrm{odd} \ . \end{cases}
\end{equation}
The situation is summarized in Fig.\ref{fig:momentum measurement zero-energy}. Perhaps surprisingly, the most probable momentum eigenvalue is now $\pm \pi/L$ rather than $0$ as one might have expected for a zero-energy state. In fact momentum zero is completely excluded. One may again explicitly compute the expectation value of $p_R$ and its square from eq.\eqref{zero-energy measurement},
\begin{equation}
    \expval{p_R} = 0 \ , \quad \quad \expval{p_R^2} = \infty \ .
\end{equation}
In contrast to the previous case, now $\expval{p_R^2}$ is infinite, and this is also true for any energy eigenstate with Neumann boundary conditions \cite{alHashimiWieseAltMomentum}. In fact, using the same spectral decomposition technique as in Section \ref{sec:momentum expval}, one may show that $\expval{p_R^2}$ is infinite on any physical, finite-energy state unless the state satisfies Dirichlet boundary conditions. This phenomenon is to be expected on physical grounds, as any momentum measurement will transfer an infinite amount of energy to the particle.

\section{Ehrenfest Theorem for a Particle in a Box}\label{sec:ehrenfest}

The Ehrenfest theorem \cite{Ehrenfest1927} is the statement that the classical relations between position, momentum, and the potential hold for the respective quantum operators within appropriately placed expectation values:
\begin{align}
    &m \dv{\expval{x}}{t} = \expval{p} \ , \label{first ehrenfest theorem} \\
    &\dv{\expval{p}}{t} = -\expval{V'(x)} \label{second ehrenfest theorem} \ ,
\end{align}
where $V'(x) = \dv{V(x)}{x}$. For a particle on the entire real line with appropriate boundary conditions, the proof of eqs.\eqref{first ehrenfest theorem} and \eqref{second ehrenfest theorem} is an elementary exercise. However, as we have seen, on a finite interval the operator $-i \p_x$ does not meaningfully represent the momentum anymore and, as such, it is not clear whether the Ehrenfest theorem holds or what is its meaning. That the Ehrenfest theorem fails for a particle on a finite interval was already noticed in the literature in connection to several topics including the domains of operators \cite{EhrenfestHill, EhrenfestAlonsoVicenzoGonzalezDiaz}, Bohmian mechanics \cite{EhrenfestBohm}, and the appearance of anomalies in quantum field theories \cite{EhrenfestAnomalies}. 

In the present section we show that the position-momentum Ehrenfest theorem holds with the new momentum concept $p_R$:
\begin{equation}
    m \dv{\expval{x}}{t} = \expval{p_R} \ .
    \label{new ehrenfest theorem}
\end{equation}
As we will show later, the statement \eqref{new ehrenfest theorem} is true for any choice of self-adjoint extension parameters for both $p_R$ and $H$, and thus for any choice of boundary conditions that make $p_R$ and $H$ self-adjoint. Moreover, we will show that
\begin{equation}
    \dv{\expval{p_R}}{t} = -\expval{V'(x)} + \expval{F_B} \ ,
    \label{new second ehrenfest theorem}
\end{equation}
where $\expval{F_B}$ is the expectation value of a force localized at the boundary. In fact, the existence of a boundary force for the particle in a box had been noticed in previous works \cite{EhrenfestAlonsoVicenzoGonzalezDiaz, Rokhsar1996}. We will provide an explicit expression for the boundary force in Section \ref{sec:second ehrenfest proof}.

The significance of these results rests first of all on the confirmation that the Ehrenfest theorem, a general result in quantum mechanics, holds also for the particle in a box, once one uses the appropriate momentum concept. Moreover, since the Ehrenfest theorem is valid for the new momentum operator $p_R$ but not for the usual momentum $-i \p_x$, unless the state satisfies Dirichlet boundary conditions, this reinforces our arguments that $p_R$ provides an appropriate notion of momentum for the particle in a box.

\subsection{The relation between $-i\p_x$ and $p_R$}\label{sec:momentum expval}

In this section we prove an intermediate result that will be of crucial importance to establish the Ehrenfest theorem. In fact, we will see that for physical states with $\Psi_e=\Psi_o$, 
\begin{equation}
    \expval{-i \p_x} = \expval{p_R} + i \expval{p_I} \ .
\end{equation}
The proof of this result requires knowledge of the quantized eigenvalues of $p_R$ eq.\eqref{pR eigenvalues}, that is $k_n = \frac{\pi}{L}n + \frac{\theta}{L}$ for $n \in \Z$ and some $\theta \in \R$. Consider now a physical state as defined in Section \ref{sec:new hamiltonian},
\begin{equation}
    \langle x \lvert \Psi \rangle = \frac{1}{\sqrt{2}} \begin{pmatrix} \Psi(x) \\ \Psi(x) \end{pmatrix} \ ,
    \label{finite-energy state}
\end{equation}
so that $\Psi(x)$ is normalized. As we noted in Section \ref{sec:new hamiltonian}, $\langle x \lvert \Psi \rangle$ does not belong to the domain of $p_R$ and therefore we cannot compute $p_R \ket{\Psi}$ by simply applying $p_R = -i \sigma_x \p_x$ to $\langle x \lvert \Psi \rangle$ in the form of eq.\eqref{finite-energy state}. However, since $p_R$ is self-adjoint, its eigenfunctions form a basis for the entire Hilbert space. Thus we may write
\begin{equation}
    \ket{\Psi} = \sum_{k} \langle \phi_k \lvert \Psi \rangle \ket{\phi_k} \ .
    \label{spectral decomposition}
\end{equation}
where the sum extends over the eigenvalues $k$ of $p_R$ and the $\ket{\phi_k}$ are the corresponding eigenstates. It is important to note that the spectral decomposition eq.\eqref{spectral decomposition} does not converge pointwise, but only in norm. The lack of convergence is evident at the endpoints, whereby the left-hand side of eq.\eqref{spectral decomposition} has $\Psi_e=\Psi_o$, while the $\phi_k$, given in eq.\eqref{pR eigenfunctions normalized}, satisfy the boundary conditions eq.\eqref{momentumbc} which make $p_R$ self-adjoint, that is $\phi_{k,o}\pqty{\pm \tfrac{L}{2}} = \lambda_{\pm} \phi_{k,e}\pqty{\pm \tfrac{L}{2}}$ with $\lambda_{\pm}$ purely imaginary, and thus the whole right-hand side will also satisfy the same boundary conditions. That the spectral decomposition of a self-adjoint operator may not converge at isolated points is not at all peculiar to this situation, and is actually the generic behavior \cite{Ree75}. The simplest example of this phenomenon is the Fourier series of the sawtooth wave, which does not converge at the endpoints of the interval. In this case, however, we are interested in computing expectation values, for which the lack of convergence at isolated points does not present an issue.

We may now proceed with the calculation of $\expval{-i \p_x}$. Using the spectral decomposition eq.\eqref{spectral decomposition} we find
\begin{equation}
    \expval{-i\p_x} \equiv \bra{\Psi}\pqty{-i\p_x}\ket{\Psi} = \sum_k \langle \Psi \lvert \phi_k \rangle \bra{\phi_k}\pqty{-i\p_x}\ket{\Psi} \ .
    \label{expval p spectral decomposition}
\end{equation}
The two-component analog of the usual momentum operator $-i \p_x$ is simply $-i \mathbbm{1} \p_x$. We can first compute $\bra{\phi_k}\pqty{-i\p_x}\ket{\Psi}$. One integration by parts leads to 
\begin{align}
    \bra{\phi_k}\pqty{-i\p_x}\ket{\Psi} &= -i\frac{1}{\sqrt{2L}} \int_{-L/2}^{L/2} dx\, e^{-ikx} \partial_x \Psi(x)= \nonumber \\
    &=-i\frac{1}{\sqrt{2L}} e^{-ikx} \Psi(x) \lvert_{-L/2}^{L/2} -i\frac{1}{\sqrt{2L}} \int_{-L/2}^{L/2} dx\, ik\,e^{-ikx} \Psi(x)= \nonumber \\
    &=-i\frac{1}{\sqrt{2L}} \bqty{e^{-ikL/2} \Psi\pqty{\tfrac{L}{2}}-e^{ikL/2} \Psi\pqty{-\tfrac{L}{2}}} + k \langle \phi_k \lvert \Psi \rangle \ .
\end{align}
We now plug this back into eq.\eqref{expval p spectral decomposition}, which leads to
\begin{align}
    \expval{-i\p_x} &= \sum_k k \langle \Psi \lvert \phi_k \rangle \langle \phi_k \lvert \Psi \rangle -\frac{i}{\sqrt{2L}} \sum_k \langle \Psi \lvert \phi_k \rangle \bqty{e^{-ikL/2} \Psi\pqty{\tfrac{L}{2}}-e^{ikL/2} \Psi\pqty{-\tfrac{L}{2}}}= \nonumber\\
    &=\expval{p_R} - \frac{i}{2L} \sum_k \int_{-L/2}^{L/2}dx\,\Psi^*(x) \bqty{\Psi\pqty{\tfrac{L}{2}} e^{ik(x-L/2)}-\Psi\pqty{-\tfrac{L}{2}} e^{ik(x+L/2)}} \label{complicated equation} \ .
\end{align}
Now, using the Poisson summation formula, one finds in this case that $\sum_k e^{ik(x-L/2)} = 2L \delta(x-L/2)$. Since the $\delta$-function is located at the boundary of the integration region, however, in this case it must be evaluated as $\int_{-L/2}^{+L/2} f(x) \delta(x - L/2) = \frac12 f\pqty{\tfrac{L}{2}}$, i.e.\ with an extra factor of $1/2$ compared to the usual prescription. This result then leads to eq.\eqref{final result expval p}.

To place this prescription on firmer ground we will give a slightly more technical argument for the same result. To this end, we will need the Poisson summation formula for a possibly discontinuous function \cite{Apostol,Trigub}, which states that
\begin{equation}
    \sum_{m \in \Z}\frac12 \bqty{f(m^-)+ f(m^+)}=\sum_{n \in \Z}\int_{-\infty}^{+\infty} f(t) \exp{\pqty{-2\pi i n t}} dt \ .
    \label{poisson summation formula}
\end{equation}
Let us consider the first sum over $k$ in eq.\eqref{complicated equation}. Using the explicit form $k = \frac{n \pi}{L} + \frac{\theta}{L}$ from eq.\eqref{pR eigenvalues} and the substitution $y = (x-L/2)/(2L)$  we obtain
\begin{align}
    \sum_k \int_{-L/2}^{L/2}dx\,\Psi^*(x) e^{ik(x-L/2)} &=\sum_{n \in \Z} \int_{-L/2}^{L/2}dx\,\Psi^*(x) e^{i \theta (x-L/2)/L} e^{i \pi n (x-L/2)/L} = \nonumber \\
    &=2L\sum_{n \in \Z} \int_{-1/2}^{0} dy\,\Psi^*(2Ly+L/2) e^{2i \theta y} e^{2 i \pi n y} \ .
\end{align}
We may now apply the Poisson summation formula eq.\eqref{poisson summation formula} with $f(y) = \Psi^*(2Ly+L/2) \exp{\pqty{2i \theta y}}$. Since the wave function $\Psi(x)$ is non-zero only for $-\tfrac{L}{2} \leq x \leq \tfrac{L}{2}$, the only non-zero contribution on the left-hand side of eq.\eqref{poisson summation formula} comes from the $m=0$ term. However, since $\Psi(x)$ is zero outside $-\tfrac{L}{2} \leq x \leq \tfrac{L}{2}$ in this case $f(0^+)=0$ and $f(0^-)=\Psi^*\pqty{\tfrac{L}{2}}$, leading to
\begin{equation}
    \sum_k \int_{-L/2}^{L/2}dx\,\Psi^*(x) e^{ik(x-L/2)} = 2L \frac12 \bqty{\Psi^*\pqty{\tfrac{L}{2}}+0} = L \Psi^*\pqty{\tfrac{L}{2}} \ .
\end{equation}
Then substituting back into eq.\eqref{complicated equation} and performing a similar calculation for the other sum, one obtains
\begin{equation}
    \expval{-i\p_x} = \expval{p_R} -\frac{i}{2} \bqty{\abs{\Psi(\tfrac{L}{2})}^2 -\abs{\Psi(-\tfrac{L}{2})}^2}= \expval{p_R} + i \expval{p_I} \ .
    \label{final result expval p}
\end{equation}
In going to the final expression we used eq.\eqref{pI matrix elements} for $\expval{p_I}$. Interestingly, since $\expval{p_R}$ is the real part of $\expval{-i\p_x}$, which has no notion of the self-adjoint extension parameters $\lambda_{\pm}$ of $p_R$, we see that actually also $\expval{p_R}$ is independent of $\lambda_{\pm}$. In particular we find that $\expval{-i\p_x} \neq \expval{p_R}$ whenever the probability density $\abs{\Psi(x)}^2$ is different at the two boundary endpoints, i.e.\ $\expval{p_I} \neq 0$. For Dirichlet boundary conditions, for example, $\expval{p_I} = 0$ and thus $\expval{-i\p_x} = \expval{p_R}$, so the Ehrenfest theorem is valid also for the usual momentum operator. However, for general Robin boundary conditions, $\expval{p_I}\neq 0$, so, as we will see in the next section, the Ehrenfest theorem is valid for the new momentum $p_R$, but not for the usual momentum $-i\p_x$.

\subsection{Proof of the position-momentum Ehrenfest theorem}\label{sec:ehrenfest proof}

We now proceed to prove the position-momentum Ehrenfest theorem eq.\eqref{new ehrenfest theorem}. Consider a physical state $\ket{\Psi}$. First we differentiate $\expval{x} = \bra{\Psi} x \ket{\Psi}$ to obtain
\begin{equation}
    \dv{}{t} \expval{x} = i\bqty{ \bra{H \Psi} x \ket{\Psi}-\bra{\Psi} x \ket{H \Psi}} \ .
    \label{derivative of x}
\end{equation}
where we used the Schr{\"o}dinger equation and the self-adjointness of $H$. Using the explicit form $H=-\frac{1}{2m} \p_x^2 +V(x)$, and integrating by parts twice we obtain
\begin{equation}
    \bra{H \Psi} x \ket{\Psi}=\bra{\Psi} x \ket{H \Psi}+\frac{1}{m} \expval{-\p_x} -\frac{1}{2m} \bqty{\pqty{\p_x \Psi^*}x\Psi - \Psi^* \pqty{\Psi +x \p_x\Psi} }_{-L/2}^{L/2} \ .
\end{equation}
The boundary term can be simplified using the Robin boundary conditions eq.\eqref{Robinbc}. The result turns out to be independent of the self-adjoint extension parameters $\gamma_{\pm}$ of the Hamiltonian, and one in fact finds
\begin{equation}
    \bqty{\pqty{\p_x \Psi^*}x\Psi - \Psi^* \pqty{\Psi +x \p_x\Psi} }_{-L/2}^{L/2} = \abs{\Psi\pqty{-\tfrac{L}{2}}}^2-\abs{\Psi\pqty{\tfrac{L}{2}}}^2=2 \expval{p_I} \ .
\end{equation}
Putting everything together, we therefore obtain
\begin{align}
    m\dv{}{t} \expval{x} &= i m\bqty{ \bra{H \Psi} x \ket{\Psi}-\bra{\Psi} x \ket{H \Psi}}= \nonumber \\
    &=\expval{-i\p_x}-i\expval{p_I} =\expval{p_R} \ ,
\end{align}
which completes the proof of the Ehrenfest theorem for any choice of self-adjoint extension parameters $\lambda_{\pm}$ and $\gamma_{\pm}$.

As already remarked, for the textbook case of Dirichlet boundary conditions,  the Ehrenfest theorem holds for both $p_R$ and the usual momentum operator $-i\p_x$. However, for general Robin boundary conditions the theorem only holds for the new momentum concept $p_R$ but not for the usual momentum $-i\p_x$. This reinforces our arguments that the new momentum concept is the appropriate physical momentum operator. 

It is worthwhile to understand why the usual infinite-volume proof of the Ehrenfest theorem fails in the finite interval. One would again start by differentiating $\expval{x}$ as in eq.\eqref{derivative of x}. Using the self-adjointness of $H$ one would then turn the right-hand side into the expectation value of the commutator $\expval{[H,x]}$ which can be formally computed by noting that $H=p^2/2m+V(x)$ and using the canonical commutation relations. Several difficulties arise with this approach in the case of a finite interval. In particular the calculation of the commutator term $H x$ is rather subtle, as, even though a wave function $\Psi(x)$ may be in $D(H)$, the combination $x\Psi(x)$ does not generally satisfy the same boundary conditions as $\Psi(x)$, and therefore there is no guarantee that $x\Psi(x) \in D(H)$ even though $\Psi(x)$ may be in $D(H)$. The same issue with domains would arise in the calculation of the commutator between $x$ and $p$. We will make further remarks on the domain of commutators in Section \ref{sec:commutator}.

\subsection{Proof of the momentum-potential Ehrenfest theorem}\label{sec:second ehrenfest proof}

We now prove the momentum-potential Ehrenfest theorem eq.\eqref{new second ehrenfest theorem}. The final expression, eq.\eqref{second ehrenfest theorem complete}, also gives an explicit form for the force at the boundary. To perform the calculation, we again consider a physical state $\ket{\Psi}$ and differentiate $\expval{p_R} = \bra{\Psi} p_R \ket{\Psi}$ to obtain
\begin{align}
    \dv{}{t} \expval{p_R} &= i\bqty{ \bra{H \Psi} p_R \ket{\Psi}-\bra{\Psi} p_R \ket{H \Psi}}=\label{derivative of pR} \\
    &= -\frac{i}{2m}\bqty{ \bra{\p_x^2 \Psi} p_R \ket{\Psi}-\bra{\Psi} p_R \ket{\p_x^2 \Psi}} + i\bqty{ \bra{V \Psi} p_R \ket{\Psi}-\bra{\Psi} p_R \ket{V \Psi}} \ . \nonumber
\end{align}
One then again inserts the spectral decomposition, $p_R = \sum_k k \ket{\phi_k}\bra{\phi_k}$, in each of the above terms. The calculation proceeds much like the one in Section \ref{sec:momentum expval}, with the main difference that now the sums over $k$ include an extra factor of $k$. However we note that
\begin{equation}
    \sum_k ik \exp{\pqty{ik(x-y)}} = \p_x \sum_k \exp{\pqty{ik(x-y)}} = -\p_y \sum_k \exp{\pqty{ik(x-y)}} \ .
    \label{sum extra k}
\end{equation}
so that this only entails an additional integration by parts. Since $\Psi$ lies in the domain of the Hamiltonian, it may be assumed to be twice differentiable but no more. Thus one will use either the derivative with respect to $x$ or $y$ in eq.\eqref{sum extra k} so that at most two derivatives act on $\Psi$. We will not repeat the whole calculation using the Poisson summation formula, as it follows the same steps as in Section \ref{sec:momentum expval}, but we note that for a general function $f(x,y)$ which is non-zero in the region $-L/2 \leq x,y \leq L/2$, the following holds:
\begin{align}
    \label{general double integral}
    \int\int \frac{dx\,dy}{2L} \sum_k ik e^{ik(x-y)} f(x,y) &= \frac12 f\pqty{\tfrac{L}{2},\tfrac{L}{2}} - \frac12 f\pqty{-\tfrac{L}{2},-\tfrac{L}{2}} - \int_{-L/2}^{L/2}dx\,\p_1 f(x,x) \nonumber \\
    &= -\frac12 f\pqty{\tfrac{L}{2},\tfrac{L}{2}} + \frac12 f\pqty{-\tfrac{L}{2},-\tfrac{L}{2}} + \int_{-L/2}^{L/2}dx\,\p_2 f(x,x) \ . 
\end{align}
where $\partial_1$ and $\partial_2$ are the partial derivatives with respect to the first and second argument respectively, while the first integral is taken in the whole region $-L/2 \leq x,y \leq L/2$. These two expressions are obtained from choosing either the derivative with respect to $x$ or $y$ in eq.\eqref{sum extra k}. The equality between the first and second line of eq.\eqref{general double integral} follows because $\p_1 f(x,x)+\p_2 f(x,x) = \dv{}{x} f(x,x)$ is a total derivative. With these preliminaries, a tedious but straightforward calculation leads to
\begin{equation}
    \label{second ehrenfest theorem complete}
    \dv{\expval{p_R}}{t} = -\expval{V'(x)} + \frac{1}{2m} \bqty{\Re{\pqty{\Psi^{\prime\prime}\Psi^*}} - \abs{\Psi'}^2}_{-L/2}^{L/2}  \ .
\end{equation}
The first term on the right-hand side of eq.\eqref{second ehrenfest theorem complete} is the usual term that is also present in the infinite volume, while in this case we also find a second term which is localized at the boundaries of the finite interval. The boundary term in eq.\eqref{second ehrenfest theorem complete} may also be rewritten so that one has
\begin{equation}
    \label{second ehrenfest theorem complete other form}
    \dv{\expval{p_R}}{t} = -\expval{V'(x)} + \frac{1}{2m} \bqty{\frac12 \dv{^2}{x^2} \abs{\Psi}^2 - 2\abs{\Psi'}^2}_{-L/2}^{L/2}  \ .
\end{equation}
Using the Robin boundary conditions eq.\eqref{Robinbc}, one finds
\begin{equation}
    \abs{\Psi'\pqty{\pm \tfrac{L}{2}}}^2 = \mp \frac12\gamma_{\pm} \Psi'\pqty{\pm \tfrac{L}{2}}^* \Psi\pqty{\pm \tfrac{L}{2}} \mp \frac12 \gamma_{\pm} \Psi\pqty{\pm \tfrac{L}{2}}^* \Psi'\pqty{\pm \tfrac{L}{2}} = \mp\frac12 \gamma_{\pm} \dv{}{x} \abs{\Psi}^2 \bigg\lvert_{\pm\tfrac{L}{2}} \ ,
\end{equation}
so that the whole second term on the right-hand side of eq.\eqref{second ehrenfest theorem complete} may be interpreted as the expectation value of a force operator $F_B$ localized at the two boundaries of the finite interval and given by the expression
\begin{align}
    \label{boundary force operator}
    F_B &= \frac{1}{2m} \lim_{\epsilon \to 0} \bigg[\frac12 \delta^{\prime\prime} (x-\tfrac{L}{2}+\epsilon) - \gamma_{+} \delta'(x-\tfrac{L}{2}+\epsilon)+\bigg. \nonumber \\
    &\quad\quad\quad\quad\quad - \frac12 \delta^{\prime\prime} (x+\tfrac{L}{2}-\epsilon) + \gamma_{-} \delta'(x+\tfrac{L}{2}-\epsilon)\bigg.\bigg] \ ,
\end{align}
where the limit is imposed so that the $\delta$-functions are computed strictly within the finite interval. The force $F_B$ is such that $\expval{F_B}$ reproduces the second term on the right-hand side of eq.\eqref{second ehrenfest theorem complete} or, equivalently, eq.\eqref{second ehrenfest theorem complete other form}. It is important to emphasize that the expression of the boundary force has been systematically derived rather than assumed, and it is interesting to note that it contains derivatives of $\delta$-functions rather than simply $\delta$-functions as one might have guessed \cite{Rokhsar1996}.

The force eq.\eqref{boundary force operator} may also be seen as arising from a potential $V_B$ localized at the two boundaries of the finite interval and given by
\begin{align}
    \label{boundary potential operator}
    V_B &= \frac{1}{2m} \lim_{\epsilon \to 0} \bigg[-\frac12 \delta' (x-\tfrac{L}{2}+\epsilon) + \gamma_{+} \delta(x-\tfrac{L}{2}+\epsilon)+\bigg. \nonumber \\
    &\quad\quad\quad\quad\quad\quad + \frac12 \delta' (x+\tfrac{L}{2}-\epsilon) - \gamma_{-} \delta(x+\tfrac{L}{2}-\epsilon)\bigg.\bigg] \ ,
\end{align}
so that $F_B = -V'_B$. The two terms involving the self-adjoint extension parameters $\gamma_\pm$ may be intuitively understood as giving rise to the Robin boundary conditions. In fact, consider the problem of a quantum particle in a $\delta$-function potential of strength $\gamma_+/2m$ located at $x=L/2$. Then integrating the Schr\"odinger equation for the wave function $\Psi$ in a small interval surrounding $x=L/2$ gives rise to the boundary condition
\begin{equation}
    \label{boundary potential interpretation}
    -\frac{1}{2m} \lim_{\epsilon \to 0} \pqty{\Psi'\pqty{\tfrac{L}{2}+\epsilon}-\Psi'\pqty{\tfrac{L}{2}-\epsilon}} + \frac{\gamma_+}{2m} \Psi\pqty{\tfrac{L}{2}} = 0  \ .
\end{equation}
Since the wave function $\Psi(x)$ is strictly zero for $x > L/2$, then eq.\eqref{boundary potential interpretation} reduces to $\Psi'\pqty{\tfrac{L}{2}}+\gamma_+\Psi\pqty{\tfrac{L}{2}}=0$, which is the Robin boundary condition eq.\eqref{Robinbc}. Unfortunately, we are unable to provide an intuitive interpretation for the $\delta'$ term in the potential $V_B$.

At the end of this section we note that one may also compute the time derivative of $\expval{p_I}$, which turns out to be
\begin{equation}
    \label{time derivative pI}
    \dv{\expval{p_I}}{t} = \frac{1}{2m} \Im{\pqty{\Psi^{\prime\prime}\Psi^*}}_{-L/2}^{L/2} \ .
\end{equation}
The calculation may be performed in the same manner as the one for the time derivative of $\expval{p_R}$ but is now much easier owing to the simple form of $p_I$. We note that in fact 
\begin{equation}
    \label{long equation pI}
    \dv{\expval{p_I}}{t} = \frac{1}{2m} \Im{\pqty{\Psi^{\prime\prime}\Psi^*}}_{-L/2}^{L/2} = \frac12 \partial_x j \lvert_{-L/2}^{L/2} = - \frac12 \partial_t \rho \lvert_{-L/2}^{L/2} \ .
\end{equation}
The first equality is just eq.\eqref{time derivative pI}, obtained through the spectral decomposition of $p_R$. The second equality follows from eq.\eqref{probability current}, the definition  of the quantum mechanical probability current $j(x)$, while the third equality, where $\rho(x) = \abs{\Psi(x)}^2$ is the probability density, follows from the continuity equation for the probability. The equality of the first and last terms in eq.\eqref{long equation pI} may immediately be seen to be true from the explicit form of $\expval{p_I}$, eq.\eqref{pI matrix elements}, and the normalization of the finite-energy state eq.\eqref{finite-energy state}. Moreover, we note that adding eq.\eqref{second ehrenfest theorem complete} and eq.\eqref{time derivative pI}, one finds
\begin{equation}
    \dv{}{t}\expval{-i\p_x} = \dv{\expval{p_R}}{t} + i \dv{\expval{p_I}}{t} = -\expval{V'(x)} + \frac{1}{2m} \bqty{\Psi^{\prime\prime}\Psi^* - \abs{\Psi'}^2}_{-L/2}^{L/2} \ ,
\end{equation}
which is the same expression as given in \cite{EhrenfestAlonsoVicenzoGonzalezDiaz}.

\section{Bouncing Wave Packets for a Particle in a Box}\label{sec:bouncing}

In light of the new concept of the momentum of the particle in a box, we would now like to reconsider the problem of a wave packet spreading and bouncing off the boundaries of the box. For a free particle on the entire real axis, we consider a momentum space Gaussian wave packet with momentum centered at $k=k_c$:
\begin{align}
    &\widetilde\psi(k,0) = \sqrt{2 a \sqrt{\pi}} 
    \exp\left(- \frac{a^2}{2}(k - k_c)^2\right) \ , \nonumber \\
    &\psi(x,t) = \sqrt{\frac{a}{a(t)^2 \sqrt{\pi}}} 
    \exp\left(- \frac{1}{2 a(t)^2}(x - \tfrac{k_c t}{m})^2\right) 
    \exp\left(i k_c x - i \frac{k_c^2}{2 m} t\right) \ , \nonumber \\
    &\frac{1}{a(t)^2} = \frac{a^2 - i \tfrac{t}{m}}{a^4 + \tfrac{t^2}{m^2}} \ .
    \label{momentum space gaussian}
\end{align}
In that case, the standard momentum operator  $p = - i \partial_x$ is
self-adjoint and one obtains
\begin{equation}
\langle x \rangle(t) = \frac{k_c}{m} t \ , \quad 
\langle p \rangle(t) = k_c \ , \quad 
\Delta x(t) = \sqrt{\frac{a^2}{2} + \frac{t^2}{2 m^2 a^2}} = 
\frac{|a(t)|^2}{\sqrt{2} a} \ , \quad
\Delta p(t) = \frac{1}{\sqrt{2} a} \ .
\end{equation}
Initially, the wave packet has a minimal uncertainty product
$\Delta x(0) \Delta p(0) = \tfrac{1}{2}$, which then increases as $\Delta x(t)$ increases. In coordinate space (but not in momentum space) it is spreading with time.

In the following sections we consider the problem of wrapping a generic solution $\Psi(x,t)$, $x \in \R$ of the free Schr{\"o}dinger equation around the finite interval $\Omega = [-\frac{L}{2}, \frac{L}{2}]$, for different choices of boundary conditions. This provides a natural way of mapping a solution on the entire real axis to the finite interval $\Omega$. In each case, we consider explicitly the momentum space Gaussian of eq.\eqref{momentum space gaussian}, which exhibits interesting behavior. The code used to produce the figures is available at \cite{valentin_wyss_2021_5212924}. 
%
\subsection{Bouncing wave packet for Dirichlet boundary 
conditions}

Let us assume that $\Psi(x,t)$ is a generic wave packet that is moving along the entire real axis. We can map this to a box with Dirichlet boundary conditions by writing
\begin{equation}
    \Psi_D(x,t) = {\cal N} \sum_{n \in \Z} [\Psi(x + 2 n L,t) - 
    \Psi(- x + (2 n + 1) L,t)] \ .
    \label{general wrapping dirichlet}
\end{equation}
Here ${\cal N}$ is an appropriate normalization factor. Due to the linearity of the Schr\"odinger equation, it is clear that $\Psi_D(x,t)$ is a solution as long as $\Psi(x,t)$ is. The wave packet $\Psi_D(x,t)$ indeed obeys Dirichlet boundary conditions because
\begin{align}
    \Psi_D(\pm \tfrac{L}{2})&=
    {\cal N} \sum_{n \in \Z} [\Psi(\pm \tfrac{L}{2} + 2 n L,t) -
    \Psi(\mp \tfrac{L}{2} + (2 n + 1) L,t)] \nonumber \\
    &={\cal N} \sum_{n \in \Z} [\Psi(\pm \tfrac{L}{2} + 2 n L,t) -
    \Psi(\mp \tfrac{L}{2} - (- 2 n \mp 1) L,t)] \nonumber \\
    &={\cal N} \sum_{n \in \Z} [\Psi(\pm \tfrac{L}{2} + 2 n L,t) -
    \Psi(\pm \tfrac{L}{2} + 2 n L,t)] = 0 \ .
    \label{Dirichletbc}
\end{align} 

\begin{figure}
    \centering
    \begin{subfigure}[b]{0.45\textwidth}
        \centering
        \includegraphics[width=6.5cm]{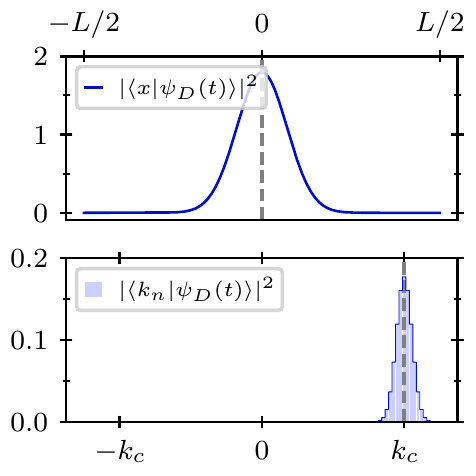}
        \caption{$t=0$}
    \end{subfigure}
    \hfill
    \begin{subfigure}[b]{0.45\textwidth}
        \centering
        \includegraphics[width=6.5cm]{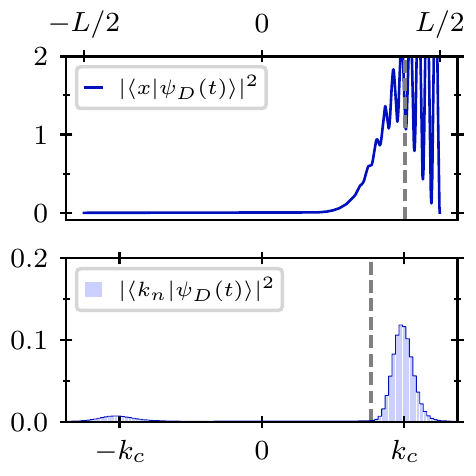}
        \caption{$t=T/400$}
    \end{subfigure}
    \hfill
    \begin{subfigure}[b]{0.45\textwidth}
        \centering
        \includegraphics[width=6.5cm]{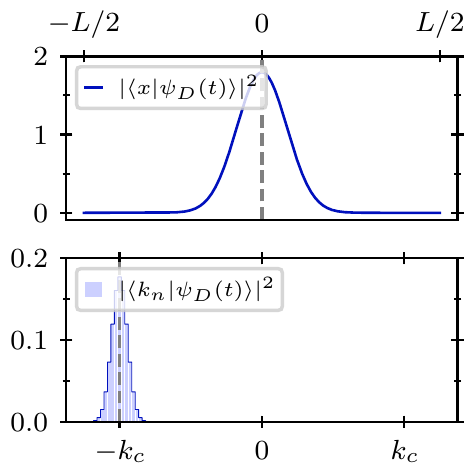}
        \caption{$t=T/2$}
    \end{subfigure}
    \hfill
    \caption{Modulus squared of the wrapped Gaussian wave packet with Dirichlet boundary conditions, as in eq.\eqref{wrapped gaussian dirichlet}, in both position (top panels) and (new) momentum space (bottom panels) at different times, where $k_c = 41\tfrac{\pi}{L}$. The dashed lines indicate the expectation value of position and (new) momentum.}
    \label{fig:gaussian bouncing evolution}
\end{figure}

It is straightforward to project the bouncing wave packet onto the energy eigenstates of the particle in the box and one obtains
\begin{align}
\langle l|\Psi_D(t)\rangle &= {\cal N} \frac{1}{\sqrt{2 L}}
[\widetilde \Psi(\tfrac{\pi}{L}(l + 1),t) + 
\widetilde \Psi(- \tfrac{\pi}{L}(l + 1),t)] \ , \quad l \ \mbox{even} \ ,
\nonumber \\
\langle l|\Psi_D(t)\rangle &= {\cal N} \frac{i}{\sqrt{2 L}}
[\widetilde \Psi(\tfrac{\pi}{L}(l + 1),t) -
\widetilde \Psi(- \tfrac{\pi}{L}(l + 1),t)] \ , \quad l \ \mbox{odd} \ .
\end{align}
Since, as we saw in Section \ref{sec:wavefunctions}, with Dirichlet boundary conditions all energy eigenvalues $E_l$ are integer-multiples of $E_0 = \tfrac{\pi^2}{2 m L^2}$, after a period 
\begin{equation}
T = \frac{2 \pi}{E_0} = \frac{4 m L^2}{\pi} \ ,
\end{equation}
the wave packet returns to its initial form. This phenomenon is known as \textit{quantum revival} \cite{QuantumRevivals, FractionalWaveFunctionRevivals}.

For the specific case of the Gaussian wave packet in eq.\eqref{momentum space gaussian}, a lengthy calculation leads to
\begin{multline}
    \psi_D(x,t) = \widetilde{\mathcal{N}} \sum_{l \in \Z} \exp{\pqty{-\frac{a^2}{2} \pqty{\frac{\pi}{L}l - k_c}^2}} \exp{\pqty{-i \frac{\pi^2 l^2}{2m L^2}t}} \times \\ 
    \times \bqty{\exp{\pqty{i\frac{\pi}{L} l x}}-(-1)^l \exp{\pqty{-i\frac{\pi}{L} l x}}} \ ,
    \label{wrapped gaussian dirichlet}
\end{multline}
where $\widetilde{\mathcal{N}}$ is a normalization factor. The time-evolution of the bouncing and spreading Gaussian wave packet $\psi_D(x,t)$ is illustrated in Fig.\ref{fig:gaussian bouncing evolution}. Initially, the packet is centered at $x = 0$ and has momentum $k_c$. While the solution on the entire real axis eq.\eqref{momentum space gaussian} keeps spreading out with time, the wrapped Gaussian never fully spreads and in fact returns to its initial shape after the revival time $T$. After half the revival time, the probability density $|\psi_D(\tfrac{T}{2})|^2 = |\psi_D(0)|^2$ returns to its original form, but the packet is then moving in the opposite direction. This is sometimes called \textit{mirror revival} \cite{QuantumRevivals, FractionalWaveFunctionRevivals}. 

\begin{figure}
    \centering
    \begin{subfigure}[b]{0.45\textwidth}
        \centering
        \includegraphics[width=6.5cm]{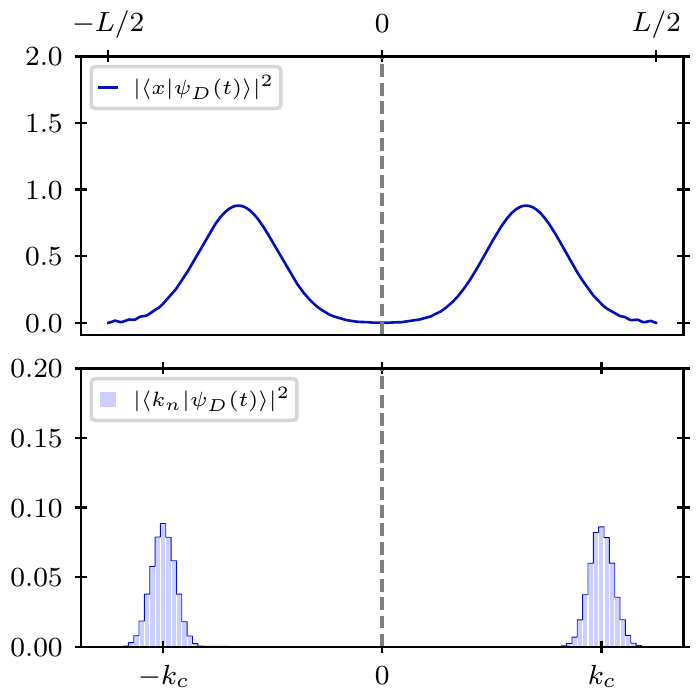}
        \caption{$t=T/4-T/625$}
    \end{subfigure}
    \hfill
    \begin{subfigure}[b]{0.45\textwidth}
        \centering
        \includegraphics[width=6.5cm]{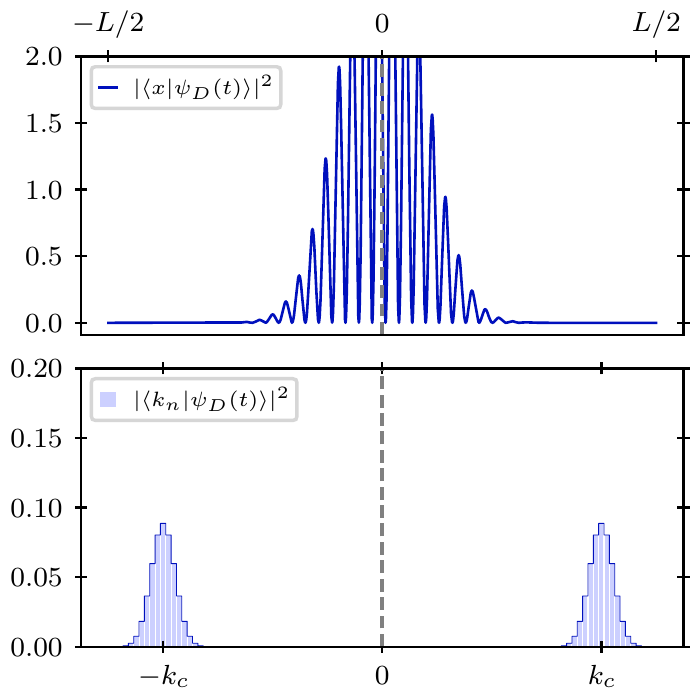}
        \caption{$t=T/4$}
    \end{subfigure}
    \hfill
    \caption{Modulus squared of the wrapped Gaussian wave packet with Dirichlet boundary conditions, as in eq.\eqref{wrapped gaussian dirichlet}, in both position and (new) momentum space shortly before and exactly at $t=T/4$, where $k_c = 41\tfrac{\pi}{L}$. The dashed lines indicate the expectation value of position and (new) momentum.}
    \label{fig:revival dirichlet T4}
\end{figure}

Something rather peculiar happens at time $t=T/4$. As shown in Fig.\ref{fig:revival dirichlet T4}, shortly before $t=T/4$ the wrapped Gaussian is visually described by the sum of two Gaussian curves with opposite momentum. This phenomenon, whereby an initial shape splits up into two or more similarly shaped curves, is sometimes called \textit{fractional revival} \cite{QuantumRevivals, FractionalWaveFunctionRevivals}. At exactly time $t=T/4$, the wave function is highly oscillatory with an overall Gaussian shape. In light of the similarity with the bouncing off the wall in Fig.\ref{fig:gaussian bouncing evolution} we may tentatively interpret this as a \say{collision} between the two Gaussians. It may also be shown analytically that in fact $\psi_D(x,T/4)$ can be obtained by wrapping onto the finite region the following momentum free particle state:
\begin{equation}
    \widetilde{\psi}(k) = \pqty{\frac{a\sqrt{\pi}}{2}}^{1/2} \bqty{(1-i) \exp{\pqty{-\frac{a^2}{2}(k-k_c)^2}}-(1+i) \exp{\pqty{-\frac{a^2}{2}(k+k_c)^2}}} \ .
\end{equation}
which, confirming the situation in Fig.\ref{fig:revival dirichlet T4}, represents the superposition of two Gaussian curves with opposite momentum and a phase shift of $\pi/2$.

\subsection{Bouncing wave packet for Neumann boundary 
conditions}

For Neumann boundary conditions we map the infinite-volume wave packet to the
box by writing
\begin{equation}
\Psi_N(x,t) = {\cal N} \sum_{n \in \Z} [\Psi(x + 2 n L,t) + 
\Psi(- x + (2 n + 1) L,t)]\ .
\end{equation}
The derivative of the wave function then obeys
\begin{equation}
\partial_x \Psi_N(x,t) = {\cal N} \sum_{n \in \Z} 
[\partial_x \Psi(x + 2 n L,t) - \partial_x \Psi(- x + (2 n + 1) L),t)] \ .
\end{equation}
Using the same manipulations as in eq.\eqref{Dirichletbc}, one concludes that
$\Psi_N(x,t)$ indeed obeys Neumann boundary conditions, i.e.\
$\partial_x \Psi_N(\pm \tfrac{L}{2},t) = 0$. 

Projecting the bouncing wave packet onto the energy eigenstates of the particle in the box with Neumann boundary conditions, one obtains
\begin{align}
\langle 0|\Psi_N(t)\rangle &= {\cal N} \frac{1}{\sqrt{L}} \widetilde \Psi(0,t)
\ , \nonumber \\
\langle l|\Psi_N(t)\rangle &= {\cal N} \frac{1}{\sqrt{2 L}}
[\widetilde \Psi(\tfrac{\pi}{L} l,t) + 
\widetilde \Psi(- \tfrac{\pi}{L} l,t)] \ , \quad l \ \mbox{even} \ ,
\nonumber \\
\langle l|\Psi_N(t)\rangle &= {\cal N} \frac{i}{\sqrt{2 L}}
[\widetilde \Psi(\tfrac{\pi}{L} l,t) -
\widetilde \Psi(- \tfrac{\pi}{L} l,t)] \ , \quad l \ \mbox{odd} \ .
\end{align}
Since, as shown in Section \ref{sec:wavefunctions}, with Neumann boundary conditions all energy eigenvalues $E_l$ are again integer-multiples of $E_1 = \tfrac{\pi^2}{2 m L^2}$, the wave packet again experiences a complete revival after the time $T = \tfrac{2 \pi}{E_1} = \tfrac{4 m L^2}{\pi}$.

For the specific case of the Gaussian wave packet in eq \eqref{momentum space gaussian}, one finds for Neumann boundary conditions,
\begin{multline}
    \psi_N(x,t) = \widetilde{\mathcal{N}} \sum_{l \in \Z} \exp{\pqty{-\frac{a^2}{2} \pqty{\frac{\pi}{L}l - k_c}^2}} \exp{\pqty{-i \frac{\pi^2 l^2}{2m L^2}t}} \times \\ \times \bqty{\exp{\pqty{i\frac{\pi}{L} l x}}+(-1)^l \exp{\pqty{-i\frac{\pi}{L} l x}}} \ ,
    \label{wrapped gaussian neumann}
\end{multline}
where $\widetilde{\mathcal{N}}$ is a normalization factor. The only difference from the wrapped Gaussian for Dirichlet boundary conditions given by eq.\eqref{wrapped gaussian dirichlet} is the plus sign in front of $(-1)^l$. 

The time-evolution of the bouncing and spreading wrapped Gaussian $\psi_N(x,t)$ follows the same lines as in the case for Dirichlet boundary conditions illustrated in Fig.\ref{fig:gaussian bouncing evolution}. The wrapped Gaussian again experiences a complete revival at time $T$, a mirror revival at time $T/2$, where the curve has the same shape but opposite momentum, and a fractional revival at time $T/4$, where the curve is described by two Gaussians with opposite momentum. The respective figures are similar to the Dirichlet case and are not displayed here.

\subsection{Bouncing wave packet for mixed 
Neumann-Dirichlet \\ boundary conditions}

Next, we consider mixed Neumann-Dirichlet boundary conditions with
$\partial_x \Psi(- \tfrac{L}{2},t) = 0$ and $\Psi(\tfrac{L}{2},t) = 0$. In this case, we map the infinite-volume wave packet to the finite box by writing
\begin{multline}
    \Psi_{ND}(x,t)={\cal N} \sum_{n \in \Z} [\Psi(x + 4 n L,t) + 
    \Psi(- x + (4 n + 1) L,t)+ \\
    -\Psi(x + (4 n + 2) L,t) - \Psi(- x + (4 n + 3) L,t)] \ .
\end{multline}
It is easy to verify that this wave function indeed obeys the appropriate boundary conditions.

\begin{figure}
    \centering
    \begin{subfigure}[b]{0.45\textwidth}
        \centering
        \includegraphics[width=6.5cm]{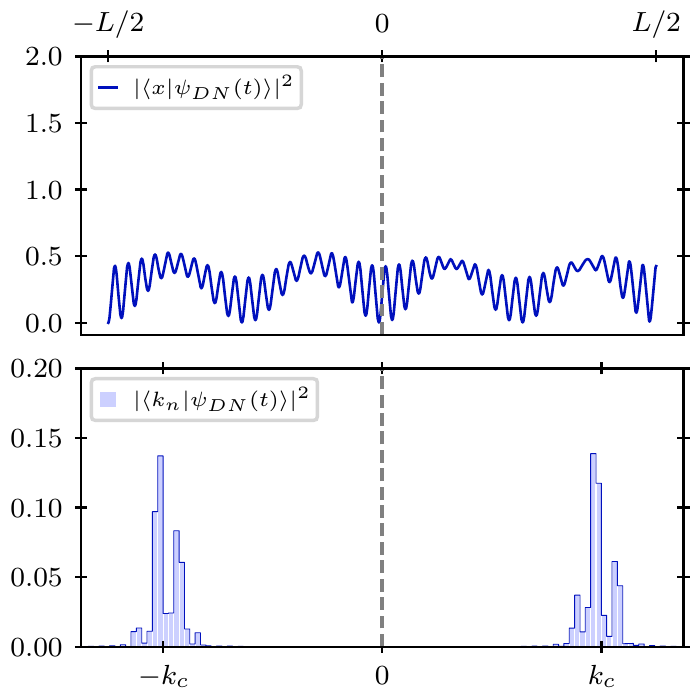}
        \caption{$t=T/8-T/1250$}
    \end{subfigure}
    \hfill
    \begin{subfigure}[b]{0.45\textwidth}
        \centering
        \includegraphics[width=6.5cm]{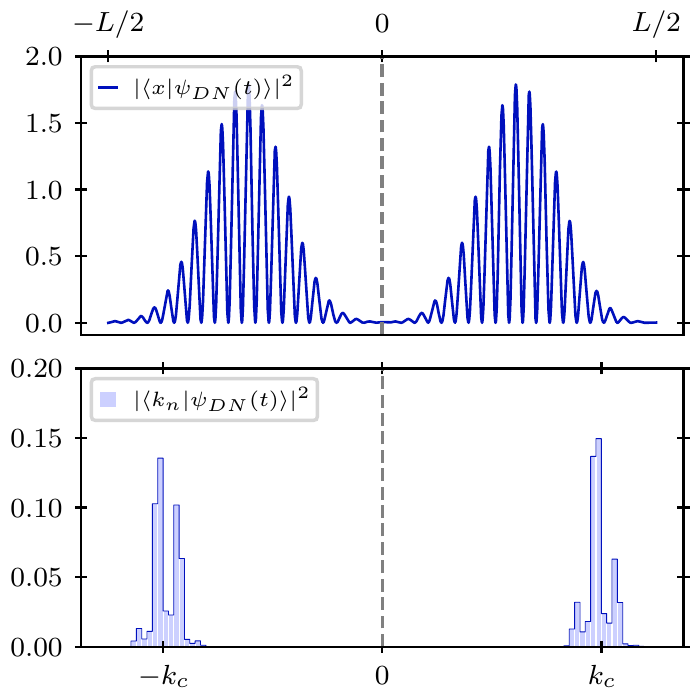}
        \caption{$t=T/8$}
    \end{subfigure}
    \hfill
    \caption{Modulus squared of the wrapped Gaussian wave packet with mixed Neumann-Dirichlet boundary conditions, in both position and (new) momentum space shortly before and exactly at $t=T/8$, where $k_c = 41\tfrac{\pi}{L}$. The dashed lines indicate the expectation value of position and (new) momentum.}
    \label{fig:mixed quadruple gaussian T8}
\end{figure}

Projecting the bouncing wave packet onto the energy eigenstates of the particle in the box with mixed Neumann-Dirichlet boundary conditions one obtains
\begin{multline}
    \label{wrapped gaussian mixed}
    \langle l|\Psi_{ND}(t)\rangle={\cal N} \frac{i}{\sqrt{2 L}}
    \left[\widetilde \Psi(\tfrac{\pi}{L} (l + \tfrac{1}{2}),t) 
    \exp\left(- i \frac{\pi}{2} (l + \tfrac{1}{2})\right)+ \right. \\
    -\left.\widetilde \Psi(- \tfrac{\pi}{L} (l + \tfrac{1}{2}),t) 
    \exp\left(i \frac{\pi}{2} (l + \tfrac{1}{2})\right)\right] \ .
\end{multline}
Unlike the two previous cases, here the energy levels are not integer multiples of a common base frequency. However, since all energy differences $E_l - E_0 = \tfrac{\pi^2}{2 m L^2} l(l+1)$ are integer-multiples of  $E_1 - E_0 = \tfrac{\pi^2}{m L^2}$, up to an irrelevant overall phase $\exp(- i E_0 t)$, the wave packet experiences a complete revival after the time 
\begin{equation}
    \frac{2 \pi}{E_1 - E_0} = \frac{2 m L^2}{\pi} = \frac{T}{2} \ ,
\end{equation}
which corresponds to half of the period $T$ for pure Dirichlet or Neumann boundary conditions. A wrapped Gaussian wave packet $\psi_{ND}(x,t)$ with mixed Neumann-Dirichlet boundary conditions thus experiences a full revival at time $T/2$. Based on experience with Dirichlet and Neumann boundary conditions, one might expect a mirror revival at half the full revival time, i.e.\ $T/4$ in this case. However, one observes a \say{double Gaussian} instead. Finally, in Fig.\ref{fig:mixed quadruple gaussian T8} we show the behavior of the wrapped Gaussian at time $T/8$, one quarter of the revival time. Shortly before time $T/8$ one observes a highly oscillatory \say{quadruple Gaussian} shape with pairs of opposite momentum, which then collide to form two highly oscillatory double Gaussians at exactly time $T/8$. Interestingly, a visually similar \say{quadruple Gaussian} is also observed at the same time $T/8$ for both Dirichlet and Neumann boundary conditions.

\section{Derivation and Interpretation of the Uncertainty Relation}\label{sec:uncertainty}

The aim of this section is to consider the uncertainty relation in light of the new momentum concept. We first compute the commutator between $x$ and $p_R$, paying careful attention to the domains of the operators involved. Since $-i\p_x$ does not qualify as the physical momentum operator, its uncertainty relation does not hold physical meaning at face value. In the subsequent sections, however, we generalize the uncertainty relation to non-Hermitean operators and then show how the uncertainty relation between $x$ and $-i\p_x$ can be interpreted as a physically meaningful inequality through the new momentum concept $\expval{p_R}$. A mathematically valid version of the Heisenberg uncertainty relation for the particle in a box was obtained in \cite{AlHashimiWieseUncertainty} but it did not admit a fully satisfactory general physical interpretation.

\subsection{Commutator between position and momentum}\label{sec:commutator}

As noted at the end of Section \ref{sec:ehrenfest proof}, one needs to be careful when computing commutators because of domain issues. In fact, if $A$ and $B$ are operators and $\Psi$ is a state, to compute $AB\Psi$ one needs not only $\Psi \in D(B)$ but also $B\Psi \in D(A)$. Therefore the domain of the commutator $[A,B]$ is given by those $\Psi \in D(A) \cap D(B)$ such that also $A\Psi \in D(B)$ and $B\Psi \in D(A)$. 

We consider this issue explicitly in the calculation of the commutator between $x$ and $p_R$. This case is simpler than the general case because on a finite interval the operator $x$ is bounded, and since it is Hermitean, it is automatically self-adjoint. More importantly, a bounded self-adjoint operator can be defined on the entire Hilbert space. This is not the case for an unbounded self-adjoint operator, which may be defined at most on a dense subset \cite{Ree75}. Since the domain of $x$ is the entire Hilbert space, no issue arises with the calculation of $x p_R \Psi$ on a state $\Psi \in D(p_R)$. On the other hand, if we want to compute $p_R x \Psi$ with $\Psi \in D(p_R)$ we need to make sure that $x\Psi \in D(p_R)$. This calculation reduces to showing that $x\Psi$ satisfies the same boundary conditions as $\Psi \in D(p_R)$, but this is easily seen to be true for any choice of self-adjoint extension parameters in the boundary conditions eq.\eqref{momentumbc}.
As such, one can easily compute the commutator on the entire domain of $p_R$:
\begin{equation}
    [x, p_R] = \begin{pmatrix}0 & i \\ i & 0\end{pmatrix} .
    \label{commutator x pR}
\end{equation}
Despite the suggestive nature of eq.\eqref{commutator x pR}, it should be emphasized that, as an operator identity, it is valid only on $D(p_R)$, and thus not on the space of interest, i.e.\ the physical subspace of finite-energy states. Nonetheless, using a spectral decomposition argument in the style of Section \ref{sec:ehrenfest proof} one can compute the expectation value of the commutator on the physical subspace:
\begin{equation}
    \expval{[x, p_R]} = i .
    \label{commutator x p_R expval}
\end{equation}
This is the same result that one would obtain by naively applying \eqref{commutator x pR} to a physical state with $\Psi_e=\Psi_o$. 
As for the imaginary part of the momentum, it is easily shown that $[x, p_I]=0$. Moreover since both $x$ and $p_I$ are bounded and Hermitean, their composition $x p_I$ is also bounded and Hermitean and thus self-adjoint and defined on the entire Hilbert space.

One may be tempted to plug the above expectation value \eqref{commutator x p_R expval} into the Heisenberg uncertainty principle to obtain an uncertainty relation between $x$ and $p_R$. However, as we saw in Section \ref{sec:momentum measurements}, the expectation value $\expval{p_R^2}$ is generally infinite, and therefore also the uncertainty in $p_R$ will be infinite. This is due to the physical fact that a momentum measurement necessarily transfers an infinite amount of energy to the particle. As such, the uncertainty relation between $x$ and $p_R$ is not meaningful.

Since the usual momentum $-i \p_x$ is not self-adjoint and thus doesn't qualify as the physical momentum operator, it is unclear what physical meaning one may assign to its uncertainty relation. However, as we will see in the next section, a generalized version of the Heisenberg uncertainty relation for $-i\p_x$ can indeed be derived and given a physical interpretation.

Incidentally, the above discussion leads to an observation that will be useful later. Since both $x$ and $p_R$ are self-adjoint, to prove the self-adjointness of the composite operator $x p_R$ one only needs to show that $x p_R$ and $p_R x$ have the same domains of definition, which is exactly what we have shown above. Hence the operators $x p_R$, $p_R x$, $\{x,p_R\}$ and $p_I x$ are all self-adjoint and thus in principle observable.

\subsection{A generalized uncertainty relation}

Although $- i \p_x$ is not a self-adjoint operator, one can use it to derive relevant inequalities, including the Heisenberg uncertainty relation,  between various quantities. The Heisenberg uncertainty relation \cite{Hei27}, which was proved rigorously by Kennard \cite{Ken27} and Weyl \cite{Wey28}, was soon generalized by Robertson \cite{Rob29} and Schr\"odinger \cite{Sch30}.
The most general form of the Heisenberg-Robertson-Schr\"odinger uncertainty relation for not necessarily Hermitean operators $A$ and $B$ follows from the minimization of the positive integral
\begin{equation}
    I = \int dx \ |\Phi(x)|^2 \geq 0, \quad \quad
    \Phi(x) = (a A + b B + \mathbbm{1}) \Psi(x),
\end{equation}
by varying $a, b \in \C$. Introducing $(\Delta A)^2 = 
\langle A^\dagger A \rangle - \langle A^\dagger \rangle \langle A \rangle$, 
$(\Delta B)^2 = \langle B^\dagger B \rangle - \langle B^\dagger 
\rangle \langle B \rangle$, where all expectation values are taken with respect to the state $\Psi$, it is straightforward to derive the remarkably simple inequality
\begin{equation}
    \Delta A \Delta B \geq \abs{ \expval{A^\dagger B} - \expval{A^\dagger}     \expval{B} } \ .
    \label{generalized uncertainty relation simple}
\end{equation}
For ease of interpretation we note that the left-hand side of eq.\eqref{generalized uncertainty relation simple} may be rewritten in a more complicated manner, so that the inequality can be equivalently stated as
\begin{align}
    \Delta A \Delta B &\geq \left[
    \left(\langle A^\dagger B + B^\dagger A \rangle - \langle A^\dagger \rangle
    \langle B \rangle - \langle B^\dagger \rangle \langle A \rangle \right)^2     \right.
    \nonumber \\
    &-\left. \left(\langle A^\dagger B - B^\dagger A \rangle - 
    \langle A^\dagger \rangle \langle B \rangle + 
    \langle B^\dagger \rangle \langle A \rangle \right)^2\right]^{1/2} \ .
    \label{generalized uncertainty relation complicated}
\end{align}
The second term on the right-hand side, which represents Robertson's extension of Heisenberg's uncertainty relation to general operators, results when one assumes that $a \in \R$. The first term, which results when one allows $a \in \C$, leads to Schr\"odinger's more stringent form of the inequality. While the minus sign in front of the second term might confuse the reader, the argument of the square root on the right-hand side is indeed the sum of two positive terms, owing to the fact that the expression in the second square is anti-Hermitean.

\subsection{Interpretation of the uncertainty relation between $x$ and $-i\p_x$}

We now apply the generalized uncertainty relation to the Hermitean (and actually self-adjoint) operator $A = A^\dagger = x$ and the non-Hermitean operator $B = - i \p_x$.
By partial integration, using Robin boundary conditions eq.\eqref{Robinbc} one obtains
\begin{align}
    \langle B^\dagger B \rangle&=\int_{-L/2}^{L/2} dx\, (-i \p_x\Psi( x))^*
     (-i \p_x\Psi(x)) \nonumber \\
    &=-\int_{-L/2}^{L/2} dx\, \Psi(x)^* \p_x^2 \Psi(x) +
    \bqty{\Psi(x)^* \p_x\Psi(x)}_{-L/2}^{L/2} 
    \nonumber \\
    &=\expval{-\p_x^2} - \bqty{\gamma_+\abs{\Psi(\tfrac{L}{2})}^2 +\gamma_-     \abs{\Psi(-\tfrac{L}{2})}^2} \ . \nonumber 
\end{align}
Here we interpret $\expval{-\p_x^2} = 2m \expval{T}$ where $T$ is the kinetic energy operator. In fact, the addition of a potential to the Hamiltonian does not change the self-adjointness conditions and the inequality that we will derive remains true in the presence of a potential. 
Since in Section \ref{sec:momentum expval} we found that $\expval{-i\p_x} = \expval{p_R} + i \expval{p_I}$, we therefore see that
\begin{equation}
    \pqty{\Delta B}^2 =  2m \expval{T} - \pqty{\gamma_+\abs{\Psi(\tfrac{L}{2})}^2 +\gamma_- \abs{\Psi(-\tfrac{L}{2})}^2} - \expval{p_R}^2 - \expval{p_I}^2 \ .
\end{equation}
The term $\expval{A^\dagger B}=\expval{x(-i\p_x)}$ on the right-hand side of the inequality \eqref{generalized uncertainty relation simple} can be computed using the orthonormal eigenstates of $p_R$ similarly to what was done in Section \ref{sec:momentum expval}. We find on a physical state $\Psi$:
\begin{align}
\bra{\phi_k} x (-i\p_x) \ket{\Psi} &= -\frac{i}{\sqrt{2L}}\int_{-L/2}^{L/2} dx \ e^{-ikx}\,x\, \p_x \Psi(x)  \\
&= -\frac{i}{\sqrt{2L}} \bqty{e^{-ikx}\,x\,\Psi(x)}_{-L/2}^{L/2}+k\bra{\phi_k} x \ket{\Psi} + i\langle\phi_k\lvert \Psi \rangle \ ,
\nonumber
\end{align}
so that therefore 
\begin{align}
\bra{\Psi} x (-i\p_x) \ket{\Psi} &= \sum_k \langle \Psi \lvert \phi_k\rangle \bra{\phi_k} x (-i\p_x) \ket{\Psi} \nonumber \\
&= \sum_k \langle \Psi \lvert \phi_k\rangle \bqty{-\frac{i}{\sqrt{2L}} \pqty{e^{-ikx}\,x\,\Psi(x)}\lvert_{-L/2}^{L/2}+k\bra{\phi_k} x \ket{\Psi} + i\langle\phi_k\lvert \Psi \rangle} \nonumber\\
&= i + \expval{p_R x}-\frac{i}{2} \frac{L}{2} \pqty{\abs{\Psi(\tfrac{L}{2})}^2+\abs{\Psi(-\tfrac{L}{2})}^2} \ .
\end{align}
Now plugging everything into the square of eq.\eqref{generalized uncertainty relation simple} and using the explicit formula eq.\eqref{pI matrix elements} for $\expval{p_I}$ one obtains the following inequality for $\expval{T}$:
\begin{align}
\label{eq:kinetic energy inequality}
2m \expval{T} &\geq \expval{p_R}^2 + \frac{1}{\pqty{\Delta x}^2} \pqty{\frac12\expval{ \{x,p_R\} }-\expval{p_R}\expval{x}}^2 \\
&+\frac{1}{4\pqty{\Delta x}^2} \bqty{1+ \pqty{\expval{x}-\frac{L}{2}}\abs{\Psi(\tfrac{L}{2})}^2 - \pqty{\expval{x}+\frac{L}{2}}\abs{\Psi(-\tfrac{L}{2})}^2 }^2 \nonumber\\
&+ \gamma_+\abs{\Psi(\tfrac{L}{2})}^2 +\gamma_- \abs{\Psi(-\tfrac{L}{2})}^2 + \frac14 \pqty{\abs{\Psi(\tfrac{L}{2})}^2 -\abs{\Psi(-\tfrac{L}{2})}^2}^2 \ ,
\nonumber
\end{align}
where $\pqty{\Delta x}^2 = \expval{x^2}-\expval{x}^2$ and the anticommutator $\{x,p_R\}$ arises when taking the absolute value squared of the right-hand side of eq.\eqref{generalized uncertainty relation simple} because $\expval{xp_R}$ is not necessarily real.

Eq.\eqref{eq:kinetic energy inequality} cannot be interpreted anymore as an ordinary uncertainty relation. However, it only contains expectation values of self-adjoint, and hence observable, operators ($T$, $x$, $x^2$, $p_R$, $\{x,p_R\}$), external parameters ($m, \gamma_{\pm}$) and values of the probability density at the endpoints, which are in principle measurable. As such, we are able to provide a meaningful physical interpretation of the uncertainty relation between $x$ and $-i\p_x$ as an inequality between measurable quantities, despite the fact that at face value the uncertainty of $-i\p_x$ has no physical meaning. This was possible only thanks to the introduction of the new momentum concept $p_R+i p_I$. Moreover, we note that eq.\eqref{eq:kinetic energy inequality} may be seen as a bound on the kinetic energy $T$, which is one of the usual applications of the ordinary Heisenberg uncertainty relation.

The inequality eq.\eqref{eq:kinetic energy inequality} is saturated by those states $\psi_S$ which satisfy $\Phi(x) = \pqty{ax + b (-i \p_x) + 1} \psi_S(x) \equiv 0$. This equation can be easily solved, leading to the unnormalized wave function
\begin{equation}
    \psi_S(x) = \exp{\bqty{-\frac{i}{b}\pqty{x+\frac12 a x^2}}} \ ,
\end{equation}
for any $a,b \in \C$ with $b \neq 0$. One then notes that 
\begin{equation}
    -\frac{1}{2m} \p_x^2 \psi_S =\frac{1}{2 m b^2} \bqty{(ax+1)^2 + i a b} \psi_S \ ,
\end{equation}
so that $\psi_S$ is an eigenstate of a Hamiltonian of the form $H=-\frac{1}{2m} \p_x^2+V(x)$ only if $a \in \R$ and $b = ic$ with $c \in \R$, in which case the corresponding potential is $V(x) = \frac{1}{2 m c^2} (ax+1)^2$. In the limit $b \to \infty$ the wave function $\psi_S$ becomes the constant zero-energy state of eqs.\eqref{neumann eigenstates} and \eqref{zero-energy states symmetric}, which thus also makes the inequality sharp. It is interesting to note that the inequality \eqref{eq:kinetic energy inequality} is satisfied for this zero energy state and, as we will soon see, also for negative energy states, in which cases one might have thought that the uncertainty relation is violated. In the special case when $a=0$ the wave function $\psi_S$ is an eigenstate of the free Hamiltonian with $\gamma_+=-\gamma_-=\gamma$, which we described in Section \ref{sec:antisymmetric hamiltonian}. The normalized wave function, eq.\eqref{eq:negative energy state}, is given by
\begin{equation}
    \psi(x) = \sqrt{\frac{\gamma}{\sinh(\gamma L)}} \exp{\pqty{-\gamma x}} \ ,
\end{equation}
and has negative energy $E=-\frac{\gamma^2}{2m}$. In fact, in this state one has
\begin{align}
    &\expval{x} = \frac{1}{2\gamma}\pqty{1-\gamma L \coth{\pqty{\gamma L}}} \ ,  \quad \quad  && \pqty{\Delta x}^2 = \frac{1}{4\gamma^2}\bqty{1+\gamma^2 L^2 \pqty{1-\coth^2{\pqty{\gamma L}}}} \ , \nonumber \\
    &\expval{T} = - \frac{\gamma^2}{2m} \ ,  && \abs{\psi{(\pm\tfrac{L}{2})}}^2 = \frac{\gamma}{\sinh(\gamma L)}\exp{\pqty{\mp \gamma L}} \ .
\end{align}
Thus in the inequality eq.\eqref{eq:kinetic energy inequality} the second line is identically zero and the third line cancels with the left-hand side. The inequality therefore reads,
\begin{align}
    0 &\geq \expval{p_R}^2 + \frac{1}{\pqty{\Delta x}^2} \pqty{\frac12\expval{     \{x,p_R\} }-\expval{p_R}\expval{x}}^2 \ ,
    \nonumber
\end{align}
which therefore implies
\begin{equation}
    \label{eq:implications}
    \expval{p_R} = \expval{\{x,p_R\}} = 0 \ .
\end{equation}
These two equations have been obtained with little effort from eq.\eqref{eq:kinetic energy inequality}, and can also be confirmed by employing the more complicated methods of Section \ref{sec:ehrenfest}. The last example involves the state
\begin{equation}
    \psi_1(x) =\sqrt{\frac{12}{L^3}} x \ ,
\end{equation}
which we already considered in eq.\eqref{zero-energy states symmetric}. This state has
\begin{equation}
    \expval{x} = \expval{T} = 0 \ , \quad \quad  \pqty{\Delta x}^2 = \frac{3}{20} L^2 \ , \quad \quad \abs{\psi_1{(\pm\tfrac{L}{2})}}^2 =\frac{3}{L} \ .
\end{equation}
Moreover, as we saw in Section \ref{sec:momentum measurements}, we have $\expval{p_R}=0$. Thus the inequality eq.\eqref{eq:kinetic energy inequality} in this case reads
\begin{equation}
\frac{16}{5} \geq \expval{ \{x,p_R\} }^2 \ ,
\end{equation}
while in fact $\expval{ \{x,p_R\} } = 0$.

\section{Conclusions}\label{sec:conclusions}

The particle in a box displays a much richer physics than previously assumed. Since the usual momentum $-i\p_x$ is not self-adjoint, we have used a new concept for the self-adjoint momentum of a particle in the finite interval $[-\tfrac{L}{2},\tfrac{L}{2}]$, first introduced in \cite{alHashimiWieseAltMomentum, alHashimiWieseHalfLine}. This requires a doubled Hilbert space of two-component wave functions, a subset of which is identified as the physical subspace that displays the same physics as the original particle in a box. However, the new momentum concept provides one with an observable momentum which may be used to perform momentum measurements and compute expectation values, which would not be possible with the usual formulation. If one insisted on using the momentum $-i\p_x$, which is self-adjoint over the entire real line $\R$, then momentum would not be quantized and a momentum measurement would catapult the particle outside the box \cite{alHashimiWieseAltMomentum, alHashimiWieseHalfLine}.

By requiring the self-adjointness of the Hamiltonian, we obtained the most general boundary conditions that are compatible with the linearity of quantum mechanics and locality of physical theories, which turn out to be Robin boundary conditions. Self-adjointness of the Hamiltonian was found to directly imply probability conservation. The resulting energy spectrum of the Hamiltonian displays several interesting features not found in the usual treatment.

The question naturally arises whether the new momentum concept satisfies several expected properties. We have shown that the Ehrenfest theorem, which is in general violated by the usual momentum $-i\p_x$, is satisfied by the new momentum concept for all choices of self-adjoint extension parameters. This reinforces our arguments that the new momentum concept is indeed the appropriate notion of momentum in a finite interval. 

Even though the usual momentum $-i\p_x$, like the new momentum $p_R$, satisfies the Heisenberg uncertainty principle, it is not immediately clear whether this relation can be given a physical interpretation. We have first shown that the uncertainty principle may be generalized to non-Hermitean operators, and thus we were able to provide an interpretation of the uncertainty relation for $-i\p_x$ as a physically meaningful inequality for the expectation value of the kinetic energy.

It would be interesting to generalize the new momentum concept to higher dimensions, following a sketch provided in \cite{alHashimiWieseAltMomentum}. Moreover, while the new momentum concept is in principle measurable, it would be interesting to construct a momentum measurement device, at least a theoretical one, along the lines first established by von Neumann \cite{VonNeumann32}.

\section*{Acknowledgments}

UJW thanks M.\!\! Al-Hashimi for his collaboration on the development of the new momentum concept in \cite{alHashimiWieseAltMomentum, alHashimiWieseHalfLine}. The research leading to these results received funding from the Schweizerischer Nationalfonds (grant agreement number 200020\_200424).

\bibliographystyle{ieeetr}
\bibliography{biblio}

\end{document}